\documentclass[pra,showpacs,twocolumn,superscriptaddress,floatfix,aps]{revtex4-2}
\usepackage{amsmath,amssymb,amsfonts,bm}
\usepackage{braket}
\usepackage{float}
\usepackage{graphicx}
\usepackage{physics}
\usepackage{mathtools}
\usepackage{hyperref}
\usepackage{comment}
\usepackage[normalem]{ulem}
\usepackage[T1]{fontenc}
\hypersetup{colorlinks,%
    linkcolor=blue,%
    citecolor=blue,%
    urlcolor=blue}
\usepackage{color}
\usepackage{tikz}
\usetikzlibrary{shapes}

\newcommand{\magenta}[1]{{\color{magenta}#1}}

\begin{document}

\title{Spinor condensate persistent currents in an atomtronic Josephson necklace}
\author{Sayan Chatterjee}
\affiliation{Department of Physics, Indian Institute of Technology Guwahati, Guwahati, Assam, 781039, India}
\author{Nalinikanta Pradhan}
\affiliation{Department of Physics, Indian Institute of Technology Guwahati, Guwahati, Assam, 781039, India}
\author{Rina Kanamoto}
\affiliation{Department of Physics, Meiji University, Kawasaki, Kanagawa 214-8571, Japan}
\author{M. Bhattacharya}
\affiliation{School of Physics and Astronomy, Rochester Institute of Technology, 84 Lomb Memorial Drive, Rochester, New York 14623, USA}
\author{Pankaj Kumar Mishra}
\affiliation{Department of Physics, Indian Institute of Technology Guwahati, Guwahati, Assam, 781039, India}

\date{\today}

\begin{abstract} 
The investigation of superflow in multi-junction Josephson circuits is currently a leading frontier of physics research,  probing the fundamental manifestations of macroscopic phase coherence and enabling applications to quantum simulation, metrology and computing. 
In this work, 
we extensively 
investigate 
the stability and dynamics of persistent currents of a spinor atomic Bose-Einstein condensate on a ring with multiple Josephson junctions. Specifically, we examine the effects of positive and negative interspecies interactions, 
co- and counter-rotation, population and Manakov asymmetry, as well as Rabi coupling on the persistent currents carried by the two components. Our analysis reveals that the presence of a second species offers multiple mechanisms for unprecedented manipulation of supercurrents on the ring, including phase-slip engineering, stability control, current-inversion switching and supercurrent pumping. Our study provides a roadmap for the engineering of persistent currents in binary ring condensates in necklace potentials, with significant implications for atomtronics, matter-wave interferometry and sensing using atomic Bose gases.
\end{abstract}

\flushbottom

\maketitle

\textit{Introduction.} The Josephson junction is a key ingredient in the development of quantum technologies~\cite{kim2025josephson}. Initially formulated for superconducting systems~\cite{kjaergaard2020superconducting}, the concept has since been extended to ultracold atomic gases~\cite{amico2021roadmap}, where it plays a pivotal role in the study of quantum phenomena. In these systems, the Josephson effect involves the tunneling of particles—whether Cooper pairs in superconductors or atoms in ultracold gases—across a weak barrier. This tunneling allows for the transfer of energy and coherence between coupled regions, facilitating the emergence of complex quantum states. Importantly, the Josephson effect has been shown to be important to atomic transport in multiply-connected geometries, such as ring traps which support quantized superfluid rotation~\cite{POLO20251}. Specifically, the recent demonstration of the stabilizing effect of multiple Josephson junctions (the ``necklace potential'') on persistent currents in scalar ring Bose-Einstein Condensates (BECs), even in the presence of dissipation, external perturbations or barrier rotation, constitutes an important, dramatic and counter-intuitive effect with substantial implications for atomtronics, simulation and metrology~\cite{Luca2024Stabilizing, nesti2026increasing, ciszak2026cooperative}.

On a parallel track, there has been a large interest in spinor (i.e. multicomponent) condensates rotating on rings without junctions.
Experiments have demonstrated spinor supercurrent decay \cite{SpinorPersistentPhysRevLett.110.025301}, 
pumping of topological supercurrents \cite{BlochoscillationsRabec2025} 
and topological charge pumping on Hall tori \cite{Halltori}. Theoretical investigations have probed fundamental issues such as persistent current \cite{RiemannTwoSpeciesPRL,SpinorPersistentStabilityPhysRevA.88.051602}, 
and rotational~\cite{SaitoPhysRevA.82.013647}
stability, and superfluid drag~\cite{NaliniPhysRevResearch.7.023051},
and applications such as atom interferometry \cite{InterferometryPhysRevLett.120.063201}
and rotation sensing~\cite{RotationPhysRevA.93.023616,RotaionPhysRevA.81.061602}. 
From these advances, it is clear that the investigation of multicomponent condensate transport in a Josephson junction array is important and timely, most immediately to clarify how the addition of barriers affects the circulation on the ring. However, this essential question is yet to be addressed.  

The present Letter represents a substantial advance by considering spinor condensate persistent currents in an atomtronic Josephson necklace. Specifically, we investigate the behavior of a two-component spinor, considering positive and negative interspecies interactions, co- and counter-rotation with various angular momenta, population and Manakov asymmetry, as well as Rabi coupling \cite{MassiveexcitationRabiPhysRevLett.128.210401}. 
We find that the presence of a second component offers unprecedented opportunities for superflow manipulation, enhancing current stabilization, allowing phase slip engineering, controlling current inversion switching, and enabling supercurrent pumping. Our analysis explores a broad parameter space for the system, offering valuable insights into the complex dynamics of multicomponent condensates and their potential applications in quantum devices and simulations.


\textit{Theoretical model and Numerical Simulation.} 
The dynamics of the two spinor condensate components formed by the $F=1$ hyperfine states $m_F = 1$ and $-1$ of the $^{23}$Na atom and coherently coupled with Rabi frequency $\Omega$~\cite{CominottiPRL128MassBEC, Kim2020Observation}, are described by the corresponding condensate wave functions $\psi_1(\phi,t)$ and $\psi_2(\phi,t)$, respectively. Their dynamics are governed by the coupled one-dimensional Gross-Pitaevskii equations~\cite{GallemiNJP2015, abad2016Persistent, Gallemi2016Coherent}

\begin{align} 
i \frac{\partial \psi_1(\phi,\tau)}{\partial \tau}
&=
\Bigl[-\frac{\partial^2}{\partial \phi^2}+ V_n(\phi)+ G_{11} |\psi_1(\phi,\tau)|^2 \Bigr. \nonumber \\ & \quad \Bigl. + G_{12} |\psi_2(\phi,\tau)|^2 \Bigr]\psi_1(\phi,\tau)
 + \Omega \, \psi_2(\phi,\tau),
\label{Eq:spinorBEC1}
\\[1em]
i \frac{\partial \psi_2(\phi,\tau)}{\partial \tau}
&=
\Bigl[-\frac{\partial^2}{\partial \phi^2}+ V_n(\phi)+ G_{22} |\psi_2(\phi,\tau)|^2 \Bigr. \nonumber \\ & \quad \Bigl. + G_{12} |\psi_1(\phi,\tau)|^2\Bigr]\psi_2(\phi,\tau)
 + \Omega \, \psi_1(\phi,\tau).
\label{Eq:spinorBEC2}
\end{align}
Justification of the one-dimensional mean field description, additional two-dimensional simulations, and the robustness of the results against thermal fluctuations obtained from the stochastic Gross-Pitaevskii equation simulations are provided in the supplementary material \cite{supplement}. 

The first terms in Eqs.~(\ref{Eq:spinorBEC1}) and  (\ref{Eq:spinorBEC2}), respectively, describe the rotational kinetic energy of the condensate confined in the ring geometry. The second term corresponds to a Josephson necklace potential formed by $n$ equally spaced Gaussian barriers~\cite{Luca2024Stabilizing,KSgan2025josephson2D,Pradhan2026Proposals} 
\begin{equation}
    V_n(\phi)  = V_0 \, \sum_{i=1}^n  e^{-2\frac{(\phi - \phi_i)^2}{\delta^2}},
\end{equation}
where $V_0$, and $\delta$ denote the barrier height and width, respectively, and $\phi_i$ specifies the barrier position. The nonlinear terms account for intra- and inter-species density-density interactions, characterized by the coupling constants $G_{11}$, $G_{22}$ and $G_{12}$, respectively, while $\Omega$ denotes the Rabi coupling strength between the two components. The ratio of the inter- and intra-species interactions is defined as $g_{12} = G_{12}/\sqrt{G_{11}G_{22}}$. To obtain the dimensionless form of the coupled Gross--Pitaevskii equations Eqs.~(\ref{Eq:spinorBEC1}) and  (\ref{Eq:spinorBEC2}), we scale the lengths by the ring radius $R$ and the energies and time by $\hbar \omega_{\beta} = \hbar^2/2 m R ^2$ and $\tau = \omega_\beta t\;$,\label{eq3} respectively, where $m$ is the atomic mass. 

The simulations were performed in two stages. First, the ground states of the binary persistent currents with winding numbers $L_{p_1}$ and $L_{p_2}$ were obtained by imaginary-time propagation of uniform condensates, $\psi_\sigma(\phi)=\sqrt{N_\sigma/2\pi}\ e^{iL_{p_\sigma}\phi}$ ($\sigma=1,2$), using Eqs.~(\ref{Eq:spinorBEC1}) and (\ref{Eq:spinorBEC2}) with a numerical time-step of $d\tau = 5 \times 10^{-5}$. These stationary states were then employed as initial conditions for real-time evolution over a duration of $0.5~\mathrm{s}$ \cite{BeattiePRL2013} with timestep $d\tau = 5 \times 10^{-7}$. The necklace potential was then generated by introducing the Josephson barriers dynamically, with their heights linearly ramped from zero to $V_0$ over $70~\mathrm{ms}$ and subsequently held constant~\cite{Mathey2016Realizing,Kunimi2019Decay,ryu2020quantum}. With this background, we now turn to the persistent current dynamics. 

\textit{Effect of inter-component density interaction.} First we consider the effect of the density-density interactions on spinor superflow with one barrier in the case where the two components have distinct winding numbers. 
\begin{figure}[!htp]
\centering
\includegraphics[width= 1\linewidth]{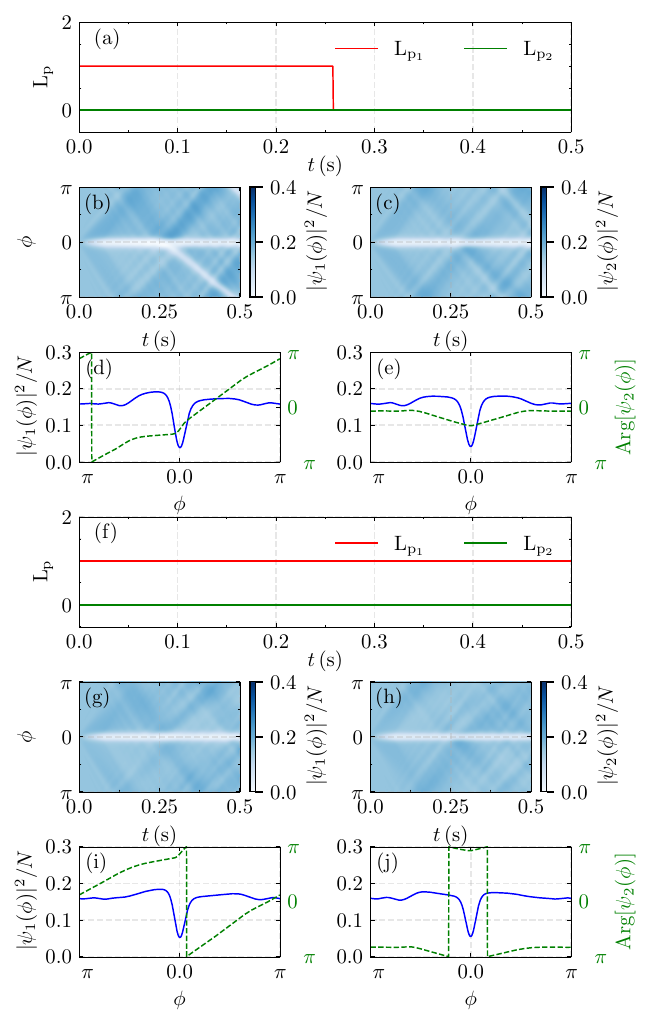} 
\caption{Dynamics of binary-persistent current states of $^{23}\mathrm{Na}$ atoms for  [(a)-(e)] $g_{12} = 0.1$ and [(f)-(j)] $g_{12} = 0.5$. [(a),(f)] Temporal evolution of the winding number of the binary-persistent current. [(b),(g)] and [(c),(h)] show the temporal evolution of the densities of $\psi_1$ and $\psi_2$, respectively.  [(d),(i)] and [(e),(j)] show the condensate density and phase profile of the first and second component, respectively, just after the barrier ramp time of $70$ ms.  The remaining parameters used are: $n_b = 1$, $L_{p_1} = 1$, $L_{p_2} = 0,\, V_0 = 2.1 \mu$, $\mu = $ chemical potential of the spinor condensate, $ \delta =  1.2 \xi$, $\xi = 1/\sqrt{\mu}$, $G_{11}=G_{22} = 2 \hbar a_s \omega_\rho/ R\hbar\omega_\beta$, $\omega_\rho = 2\pi \times 42$ Hz, $a_s = 2.5$ nm \cite{KumarPRL2021}, $\int |\psi_1(\theta)|^2 d\theta = N_1$, $\int |\psi_2(\theta)|^2 d\theta = N_2$, $N_1 = N_2 = N = 10000,\,$ and $ R = 25 \mu \mathrm{m}$.}
\label{fig:fig_nb_1_G12_0.1}
\end{figure}
As shown in Fig.~\ref{fig:fig_nb_1_G12_0.1}(a), for a small value of $g_{12}= 0.1$, the initially rotating state spontaneously decays as $L_{p1} = 1\rightarrow 0$, while the winding number of the non-rotating component $L_{p2}=0$ remains invariant. Correspondingly, the density of the second component remains uniform in time [Fig.~\ref{fig:fig_nb_1_G12_0.1} (b)]. In contrast,the density evolution of the first component  [Fig.~\ref{fig:fig_nb_1_G12_0.1} (c)] shows that the decay of $L_{p1}$ occurs by the emission of a single soliton, seen from the emergence of a new density dip near the barrier position and its motion in the ring at $t \sim 0.25$ s, the time at which $L_{p1}$ decreases in value. In this case, the two components are nearly noninteracting, and the winding number of the rotating component drops since the phase jump per barrier $\delta \phi$ is greater than the critical phase-jump $\delta \phi_c$, as also recently reported in the case of a single-component BEC in a Josephson necklace~\cite{Luca2024Stabilizing}.

\begin{figure}[!htp]
\centering
\includegraphics[width= 1\linewidth]{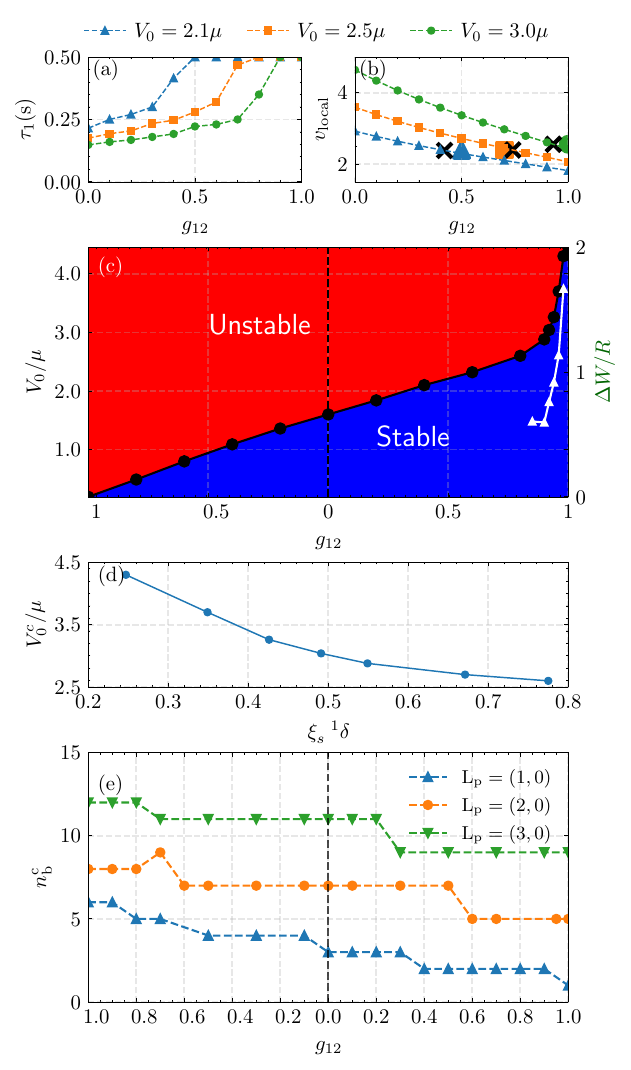} 
\caption{(a) Persistent-current lifetime $\tau_1$ as a function of the intercomponent interaction strength $g_{12}$ for three barrier heights, $V_0=2.1\mu$, $2.5\mu$, and $3.0\mu$. The lifetime of the $L_{p1}=1$ state increases monotonically with $g_{12}$. (b) Local superfluid velocity $v_{\mathrm{local}}$ of the $L_{p1}=1$ component at the barrier position, evaluated immediately after the barrier ramp, as a function of $g_{12}$. Crosses denote the critical $g_{12}$ for stability and the corresponding local velocity. Larger markers indicate the onset of the regime where the flow velocity falls below the local speed of sound 
(c) Stability phase diagram in the $V_0$-$g_{12}$ parameter space. The red region denotes parameters for which the $L_{p1}=1$ state undergoes a phase slip to the lower-winding state $L_{p1}=0$ via soliton emission, whereas the blue region denotes the topologically protected $L_{p1}=1$ state. The white curve represents the variation of the soliton's width $\Delta W$ with high values of $g_{12}$. (d) Critical barrier height $V_{0}^{c}$ as a function of the spin healing length $\xi_s$, tuned through $g_{12}$ at fixed barrier width $\delta$. (e) Critical number of barriers ($n_{b}^{c}$) required to stabilize the binary-persistent current as a function of density-density interaction strength ($g_{12}$). Here, $V_0 = 3.0 \mu$ and the other parameters are the same as in Fig.~\ref{fig:fig_nb_1_G12_0.1}.}
\label{fig:lifetime}
\end{figure}

In the same work it was shown that increasing the number of Josephson junctions provides current stability by lowering the phase jump per barrier below $\delta \phi_c$ and preventing the spontaneous release of solitons \cite{Luca2024Stabilizing}. We now demonstrate that the rotating state $L_{p1}$ can be stabilized without increasing the number of barriers, but instead by using a higher strength of interspecies density interaction $g_{12}$. The corresponding temporal evolutions of winding number and condensate density are shown in Fig.~\ref{fig:fig_nb_1_G12_0.1}[(f)-(j)]. For the higher coupling strength $g_{12} = 0.5 $, the winding number $L_{p1}$ stays constant up to the simulation time $t = 0.5$s [Fig.~\ref{fig:fig_nb_1_G12_0.1}(f)], in contrast to what is observed for lower coupling and the same number of barriers  [Fig.~\ref{fig:fig_nb_1_G12_0.1}(a)].

As $g_{12}$ is increased, phase slips are progressively delayed, and eventually suppressed, over the observation time $t=0.5$ s. This trend is quantified in Fig.~\ref{fig:lifetime} (a), where the persistent-current lifetime $\tau_{1}$ increases monotonically with $g_{12}$. The enhanced stability originates from a reduction of the local flow velocity $v_{\mathrm{local}}$  at the barriers [Fig.~\ref{fig:lifetime}(b)]~\cite{supplement}. 
At fixed barrier height $V_{0}$, this reduction results from the increase in the chemical potential of the rotating component, $\mu_{\mathrm{rot}}$. Consequently, increasing $g_{12}$ lowers the dimensionless effective barrier height, $V_{\mathrm{eff}}=V_0/\mu_{\mathrm{rot}}$. When $v_{\mathrm{local}}$ falls below the local sound velocity $v_{\mathrm{sound}}$ 
~\cite{supplement}, the persistent current satisfies the Landau stability criterion~\cite{PiazzaJPB2013,nesti2026increasing}. In contrast, at fixed $g_{12}$, increasing $V_0$ deepens the density minima at the barriers and enhances $v_{\mathrm{local}}$, thereby facilitating soliton emission and reducing the current lifetime. 

This competition between barrier strength and intercomponent interaction is summarized by the $V_0$-$g_{12}$ phase diagram of Fig.~\ref{fig:lifetime} (c). The critical barrier strength $V_{0}^{c}$ required for triggering a phase slip increases monotonically with $g_{12}$, reflecting the suppression of the local flow velocity. Strikingly, this critical strength rises sharply near the miscible-immiscible boundary, $g_{12}\simeq0.9$, resulting in a pronounced region of anomalously strong stability of the initial winding number. This behavior is a consequence of the pronounced softening of the spin mode in the spin-dominated regime. As $g_{12}\rightarrow 1$ $(G_{11}= G_{12} = G_{22})$, the spin healing length $\xi_s\propto[2n_b(G_{11}-G_{12})]^{-1/2}$~\cite{StringariPhysRevLett.116.160402,FerrariPhysRevLett.125.030401} diverges, so that the dimensionless barrier width $\sigma/\xi_s$ becomes vanishingly small. The barrier is therefore effectively transparent at the scale of the spin healing length, strongly suppressing phase slips and yielding an exceptionally robust binary persistent current [Fig.~\ref{fig:lifetime}(d)]. When a phase slip does occur in this regime, it is accompanied by the emission of broad, spin-dominated magnetic solitons~\cite{StringariPhysRevLett.116.160402}. Their increasing spatial extent with $g_{12}$ follows the growth of $\xi_s$ [right axis of Fig.~\ref{fig:lifetime}(c)]~\cite{supplement}. 

The stabilization induced by intercomponent interactions nevertheless has a finite range. For sufficiently strong barriers, e.g., $V_0=3.0\mu$, the local flow velocity remains above the Landau critical velocity even at large $g_{12}$, and phase slips therefore persist. In this regime, stability can instead be recovered by increasing the number of barriers or reducing the initial winding number $L_{p1}$. Figure~\ref{fig:lifetime} (e) summarizes this behaviour through the critical barrier number as a function of $g_{12}$ and $L_p$, showing that stronger intercomponent interactions require substantially fewer barriers to stabilize the binary persistent current. The stabilization can be further enhanced by adjusting the population imbalance between the two components~\cite{BeattiePRL2013,supplement}.

\begin{figure}[!htb]
\centering
\includegraphics[width=1.0\linewidth]{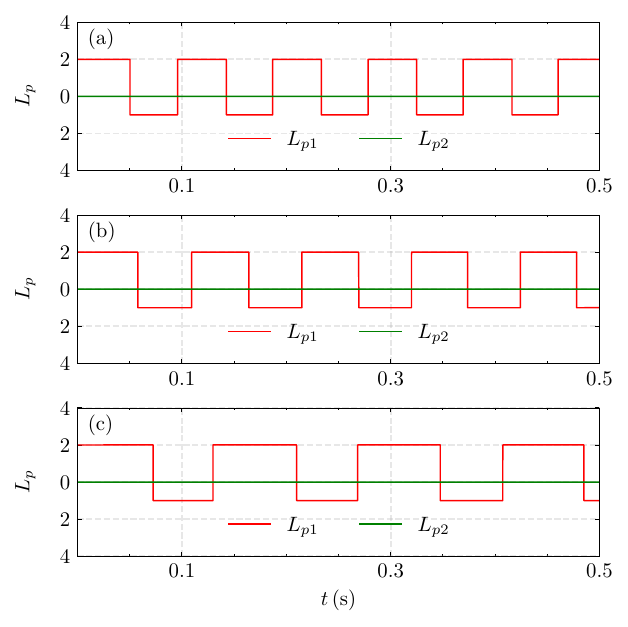}
\caption{Temporal evolution of winding numbers the spinor persistent current for (a) $g_{12} = 0.1$, (b) $g_{12} = 0.5$, and (c) $g_{12} = 0.9$.  Here $L_{p}= (2,0)$, $n_b = 3$, and $V_0 = 3.0\mu$. The other parameters are the same as in Fig. ~\ref{fig:fig_nb_1_G12_0.1}. From (a) to (c), as $g_{12}$ increases, a decrease in switching frequency of the winding number $L_{p1}$ is observed.}
\label{fig:Wind_osc}
\end{figure}

\begin{figure}[!htp]
\centering
\includegraphics[width=1.0\linewidth]{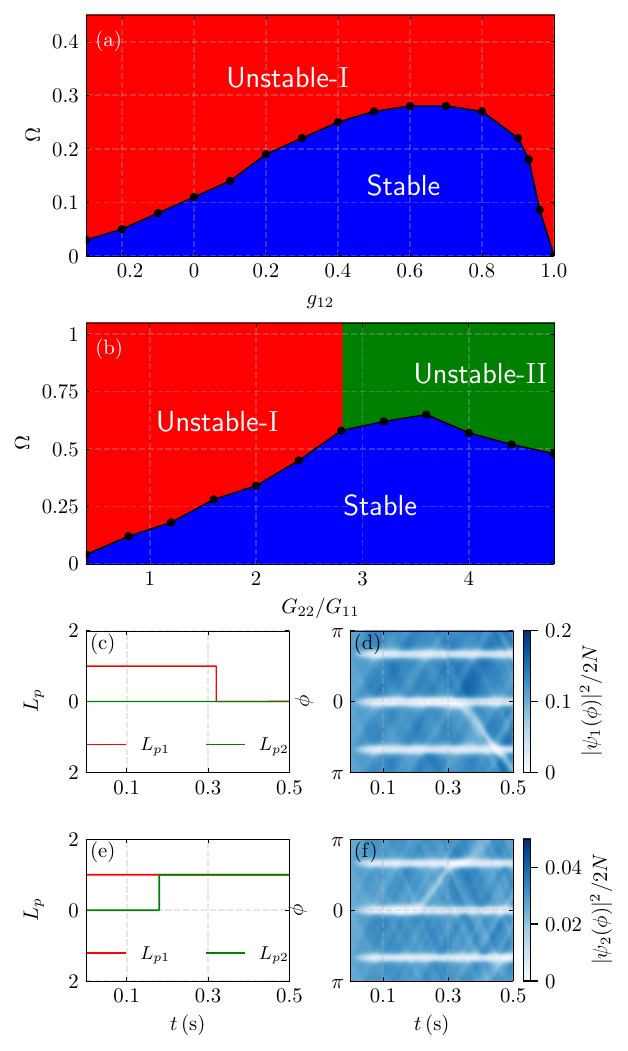} 
\caption{(a) Stability phase diagram in the $\Omega - g_{12}$ parameter space for $n_b = 3$ and $V_0 = 2.1 \mu$ for $G_{11} = G_{22}$. (b) Stability phase diagrams corresponding to persistent-current instabilities in the broken-Manakov regime ($G_{11}\neq G_{22}$) for a ring condensate with three symmetrically placed barriers ($V_0=2.1\mu$) for $g_{12}=0.1\sqrt{G_{11}G_{22}}$. Red (Unstable-I) and green (Unstable-II) regions correspond to the instability channels shown in (c,d) and (e,f), respectively. (c,d) Decay $L_p=(1,0) \rightarrow (0,0)$ via a phase slip in the first component, showing the winding-number evolution (c) and density (normalized to total number of particles) dynamics (d). Parameters are $G_{22}=2.0G_{11}$ and $\Omega=0.35$. (e,f) Instability leading to supercurrent pumping $L_p=(1,0)\rightarrow(1,1)$, showing the winding-number evolution (e) and density
dynamics (f). Here the emitted soliton propagates in a direction opposite to that in (d). Parameters are $G_{22}=4.4G_{11}$ and $\Omega=0.55$.}
\label{g22.ne.g11}
\end{figure}

\textit{Persistent current switching.} Further, we show that beyond stabilizing persistent currents, the interplay between the Josephson barriers and intercomponent interactions enables controlled switching between distinct topological states of the condensate. As discussed above, Josephson barriers can trigger phase slips that change the condensate circulation. For suitable barrier configurations, however, these phase slips become deterministic rather than irreversible, resulting in repeated transitions between quantized circulation states and periodic oscillations of the winding number~\cite{nesti2026increasing, ObiolPRR2022}.

This mechanism is illustrated in Fig.~\ref{fig:Wind_osc} through the time evolution of the winding numbers and condensate densities for different intercomponent interaction strengths $g_{12}$. The condensate is initially prepared in the state $L_p=(2,0)$ with three equally spaced Josephson barriers ($n_b=3$) of height $V_0=3.0\mu$. The system undergoes coherent switching between the states $L_{p1}$ and $L_{p1}-n_b$, i.e. with each phase slip changing the circulation by three quanta. Remarkably, we find that the switching frequency can be continuously tuned by varying $g_{12}$ [Figs.~\ref{fig:Wind_osc}(a)-(c)]. Specifically, increasing $g_{12}$ from $0.1$ to $0.9$ progressively increases the oscillation period, as stronger intercomponent interaction makes the persistent current more stable, thereby reducing the switching rate while keeping the dynamics coherent. 



 
The interplay between intercomponent interactions (tunable using magnetic fields \cite{TwoSpeciesFeshbachPRL2008}) and Josephson barriers therefore enables coherent and controllable switching between quantized persistent-current states, rather than irreversible superflow decay. 
Such controllable topological dynamics offers a promising route toward atomtronic switching devices and the manipulation of quantized supercurrents in multicomponent BECs.


\textit{Effect of Rabi coupling with Manakov symmetry  ($G_{11}=G_{22}$).}
We now consider the effect of Rabi coupling between the two components with strength $\Omega$. 
Unlike the density-mediated intercomponent interaction $g_{12}$, the Rabi term directly couples the phases of the condensate component and contributes an energy $E_r = -|\psi_1||\psi_2|\Omega \cos(\Delta\phi)$, where $\Delta\phi$ is the relative phase~\cite{PhysRevA.100.013630,PhysRevA.91.063635}. For $\Omega>0$, the energetically favored configuration is a phase-locked state with $\Delta\phi=0$, implying that initial states with unequal windings, such as $L_p=(1,0)$, are energetically frustrated. To reduce this phase mismatch, the system eliminates the circulation in the second component, making Rabi coupling effectively destabilizing in this configuration, and leading to rotational collapse. 

To further elucidate, in the case of unequal rotation $L_{p}=(1,0)$, the competition between density-density interactions (which stabilize the decay) and coherent Rabi coupling (which promotes the decay), we show the $g_{12}$-$\Omega$ phase diagram in Fig.~\ref{g22.ne.g11} (a). For weak interspecies interaction, $g_{12}\ll 1$, the two components remain nearly homogeneous, and increasing $g_{12}$ enhances the energetic cost of relative-density fluctuations associated with a phase slip. The circulating state is therefore stabilized, and a larger Rabi coupling, $\Omega$ is required to trigger the instability. As $g_{12}$ approaches the miscible-immiscible boundary, however, the role it plays changes qualitatively. The increasing interspecies repulsion now favors spatial segregation and generates pronounced density modulations that act as weak links for the superflow, thereby lowering the local energy barrier for phase slips.  Consequently, the critical $\Omega$ decreases with $g_{12}$ in this regime, in agreement with earlier reports~\cite{PhysRevA.91.063635}. 
Thus, the nonmonotonic phase boundary reflects a crossover from interaction-induced stabilization in the homogeneous regime to density-segregation-assisted phase slips near the miscible-immiscible boundary~\cite{voronova2025exciton, liu2015new}. 

\textit{Effect of Rabi coupling in the absence of Manakov symmetry ($G_{11}\neq G_{22}$).} We now consider the experimentally relevant case $G_{11}\neq G_{22}$, where the broken Manakov symmetry~\cite{Manakov:1974, Sarkar:2025} gives rise to instability pathways qualitatively distinct from those in the symmetric case. We fix $g_{12}=0.1\sqrt{G_{11}G_{22}}$, such that varying $G_{22}$ simultaneously changes the intercomponent interaction, while keeping the mixture miscible. The resulting dynamics for the initial state $L_p=(1,0)$ are shown in Fig.~\ref{g22.ne.g11}[(b)-(f)]. For small $G_{22}/G_{11}$, we observe the conventional decay $L_p=(1,0)\rightarrow(0,0)$ [Figs.~\ref{g22.ne.g11}(c),(d)], accompanied by the emission of a soliton when the local superfluid velocity exceeds the critical value. Via this rotational collapse, the Rabi energy $E_{r}$ is minimized. For sufficiently large $G_{22}/G_{11}$, remarkably, this decay channel is suppressed, and instead we observe $L_p=(1,0)\rightarrow(1,1)$ [Figs.~\ref{g22.ne.g11}(e), (f)], i.e. the initially rotating component retains its circulation, while the nonrotating component \textit{acquires} one quantum of circulation. This strikingly different behavior originates from the increasing stiffness of the second component: strong self-repulsion suppresses density modulations and makes the conventional soliton-mediated loss of circulation energetically costly.  Therefore, Rabi coupling transfers angular momentum into phase winding rather than producing a phase slip that destroys the existing circulation.  The resulting soliton emission occurs with the opposite propagation direction and accompanies the acquisition of one unit of winding by the second component, allowing the two components to become comoving while reducing both interaction and Rabi energies. Thus, breaking the Manakov symmetry transforms Rabi-induced phase slips from a circulation-loss process into a pathway for controlled circulation transfer and persistent-current pumping. 

\textit{Conclusion and future scope.} In this Letter, we have demonstrated that binary Bose-Einstein condensates in ring-shaped Josephson junction arrays provide a highly tunable platform for controlling persistent currents through the interplay of nonlinear interactions, coherent coupling, and topology. 
These results reveal a rich landscape of persistent-current dynamics that has no analog in single-component systems. Although the present analysis is restricted to one dimension, 
we have verified this through extensive two-dimensional simulations (closer to current experiments) which will be presented elsewhere, in a longer work. Our work establishes multicomponent Josephson-ring condensates as a promising platform for programmable atomtronic circuits, matter-wave interferometry and supercurrent engineering. The mechanisms identified here provide unprecedented routes for manipulating multispecies quantized supercurrents and motivate future studies of finite-temperature effects, higher-dimensional geometries, and nonequilibrium driving in multicomponent superfluids.

\textit{Acknowledgment.} M.B. thanks the Air Force Office of Scientific Research (FA9550-23-1-0259) for support. R.K. was supported by JSPS KAKENHI
(Grant No. 25K07190).
\bibliography{citation.bib} 

@article{KumarPRL2021,
  title = {{Cavity Optomechanical Sensing and Manipulation of an Atomic Persistent Current}},
  author = {Kumar, Pardeep and Biswas, Tushar and Feliz, Kristian and Kanamoto, Rina and Chang, M.-S. and Jha, Anand K. and Bhattacharya, M.},
  journal = {Phys. Rev. Lett.},
  volume = {127},
  issue = {11},
  pages = {113601},
  numpages = {7},
  year = {2021},
  month = {Sep},
  publisher = {American Physical Society},
  doi = {10.1103/PhysRevLett.127.113601},
  url = {https://link.aps.org/doi/10.1103/PhysRevLett.127.113601}
}

@article{BeattiePRL2013,
  title = {{Persistent Currents in Spinor Condensates}},
  author = {Beattie, Scott and Moulder, Stuart and Fletcher, Richard J. and Hadzibabic, Zoran},
  journal = {Phys. Rev. Lett.},
  volume = {110},
  issue = {2},
  pages = {025301},
  numpages = {5},
  year = {2013},
  month = {Jan},
  publisher = {American Physical Society},
  doi = {10.1103/PhysRevLett.110.025301},
  url = {https://link.aps.org/doi/10.1103/PhysRevLett.110.025301}
}

@article{PoloPRL2019,
  title = {{Oscillations and Decay of Superfluid Currents in a One-Dimensional {B}ose Gas on a Ring}},
  author = {Polo, Juan and Dubessy, Romain and Pedri, Paolo and Perrin, H\'el\`ene and Minguzzi, Anna},
  journal = {Phys. Rev. Lett.},
  volume = {123},
  issue = {19},
  pages = {195301},
  numpages = {6},
  year = {2019},
  month = {Nov},
  publisher = {American Physical Society},
  doi = {10.1103/PhysRevLett.123.195301},
  url = {https://link.aps.org/doi/10.1103/PhysRevLett.123.195301}
}

@article{ObiolPRR2022,
  title = {Coherent phase slips in coupled matter-wave circuits},
  author = {P\'erez-Obiol, A. and Polo, J. and Amico, L.},
  journal = {Phys. Rev. Res.},
  volume = {4},
  issue = {2},
  pages = {L022038},
  numpages = {6},
  year = {2022},
  month = {May},
  publisher = {American Physical Society},
  doi = {10.1103/PhysRevResearch.4.L022038},
  url = {https://link.aps.org/doi/10.1103/PhysRevResearch.4.L022038}
}

@article{GallemiNJP2015,
  title = {Coherent quantum phase slip in two-component bosonic atomtronic circuits},
  author = {Gallemí, A and Mateo, A Muñoz and Mayol, R. and Guilleumas, M.},
  journal = {New J. Phys.},
  volume = {18},
  issue = {},
  pages = {015003},
  numpages = {12},
  year = {2015},
  month = {Dec},
  publisher = {IOP Publishing},
  doi = {10.1088/1367-2630/18/1/015003},
  url = {https://iopscience.iop.org/article/10.1088/1367-2630/18/1/015003}
}

@article{PiazzaJPB2013,
  title = {{Critical velocity for a toroidal {B}ose–{E}instein condensate flowing
through a barrier}},
  author = {Piazza, F and Collins, L A and Smerzi, A},
  journal = {J. Phys. B: At. Mol. and Opt. Phys.},
  volume = {46},
  issue = {},
  pages = {095302},
  numpages = {7},
  year = {2013},
  month = {},
  publisher = {IOP Publishing},
  doi = {10.1088/0953-4075/46/9/095302},
  url = {https://iopscience.iop.org/article/10.1088/0953-4075/46/9/095302}
}

@article{pradhan2024cavity,
  title={Cavity optomechanical detection of persistent currents and solitons in a bosonic ring condensate},
  author={Pradhan, Nalinikanta and Kumar, Pardeep and Kanamoto, Rina and Dey, Tarak Nath and Bhattacharya, M and Mishra, Pankaj Kumar},
  journal={Physical Review Research},
  volume={6},
  number={1},
  pages={013104},
  year={2024},
  publisher={APS},
  doi = {10.1103/PhysRevResearch.6.013104},
  url = {https://link.aps.org/doi/10.1103/PhysRevResearch.6.013104}
}

@article{das2012winding,
  title={{Winding up superfluid in a torus via {B}ose {E}instein condensation}},
  author={Das, Arnab and Sabbatini, Jacopo and Zurek, Wojciech H},
  journal={Scientific reports},
  volume={2},
  number={1},
  pages={1--6},
  year={2012},
  publisher={Springer},
  doi = {https://doi.org/10.1038/srep00352},
  url = {https://www.nature.com/articles/srep00352}
}

@article{stoof2001dynamics,
  title={Dynamics of fluctuating {B}ose--{E}instein condensates},
  author={Stoof, HTC and Bijlsma, MJ},
  journal={Journal of low temperature physics},
  volume={124},
  pages={431--442},
  year={2001},
  publisher={Springer}
}

@article{abad2016Persistent,
  title = {Persistent currents in coherently coupled Bose-Einstein condensates in a ring trap},
  author = {Abad, Marta},
  journal = {Phys. Rev. A},
  volume = {93},
  issue = {3},
  pages = {033603},
  numpages = {11},
  year = {2016},
  month = {Mar},
  publisher = {American Physical Society},
  doi = {10.1103/PhysRevA.93.033603},
  url = {https://link.aps.org/doi/10.1103/PhysRevA.93.033603}
}

@article{pradhan2025AndreevBashkin,
  title = {Signature of Andreev-Bashkin superfluid drag from cavity optomechanics},
  author = {Pradhan, Nalinikanta and Kanamoto, Rina and Bhattacharya, M. and Mishra, Pankaj Kumar},
  journal = {Phys. Rev. Res.},
  volume = {7},
  issue = {2},
  pages = {023051},
  numpages = {11},
  year = {2025},
  month = {Apr},
  publisher = {American Physical Society},
  doi = {10.1103/PhysRevResearch.7.023051},
  url = {https://link.aps.org/doi/10.1103/PhysRevResearch.7.023051}
}

@article{CominottiPRL128MassBEC,
  title = {Observation of Massless and Massive Collective Excitations with {F}araday Patterns in a Two-Component Superfluid},
  author = {Cominotti, R. and Berti, A. and Farolfi, A. and Zenesini, A. and Lamporesi, G. and Carusotto, I. and Recati, A. and Ferrari, G.},
  journal = {Phys. Rev. Lett.},
  volume = {128},
  issue = {21},
  pages = {210401},
  numpages = {6},
  year = {2022},
  month = {May},
  publisher = {American Physical Society},
  doi = {10.1103/PhysRevLett.128.210401},
  url = {https://link.aps.org/doi/10.1103/PhysRevLett.128.210401}
}

@article{Luca2024Stabilizing ,
  title={Stabilizing persistent currents in an atomtronic Josephson junction necklace},
  author={Pezz{\`e}, Luca and Xhani, Klejdja and Daix, Cyprien and Grani, Nicola and Donelli, Beatrice and Scazza, Francesco and Hernandez-Rajkov, Diego and Kwon, Woo Jin and Del Pace, Giulia and Roati, Giacomo},
  journal={Nature Communications},
  volume={15},
  number={1},
  pages={4831},
  year={2024},
  publisher={Nature Publishing Group UK London}
}

@article{ryu2020quantum,
  title={Quantum interference of currents in an atomtronic SQUID},
  author={Ryu, Changhyun and Samson, EC and Boshier, Malcolm Geoffrey},
  journal={Nature communications},
  volume={11},
  number={1},
  pages={3338},
  year={2020},
  publisher={Nature Publishing Group UK London},
  url = {https://www.nature.com/articles/s41467-020-17185-6}
}

@article{voronova2025exciton,
  title={Exciton-polariton ring Josephson junction},
  author={Voronova, Nina and Grudinina, Anna and Panico, Riccardo and Trypogeorgos, Dimitris and De Giorgi, Milena and Baldwin, Kirk and Pfeiffer, Loren and Sanvitto, Daniele and Ballarini, Dario},
  journal={Nature Communications},
  volume={16},
  number={1},
  pages={466},
  year={2025},
  publisher={Nature Publishing Group UK London},
  url = {https://www.nature.com/articles/s41467-024-55119-8}
}

@article{RiemannTwoSpeciesPRL,
  title = {Mixtures of Bose Gases Confined in a Ring Potential},
  author = {Smyrnakis, J. and Bargi, S. and Kavoulakis, G. M. and Magiropoulos, M. and K\"arkk\"ainen, K. and Reimann, S. M.},
  journal = {Phys. Rev. Lett.},
  volume = {103},
  issue = {10},
  pages = {100404},
  numpages = {4},
  year = {2009},
  month = {Sep},
  publisher = {American Physical Society},
  doi = {10.1103/PhysRevLett.103.100404},
  url = {https://link.aps.org/doi/10.1103/PhysRevLett.103.100404}
}

@article{TwoSpeciesFeshbachPRL2008,
  title = {Double Species Bose-Einstein Condensate with Tunable Interspecies Interactions},
  author = {Thalhammer, G. and Barontini, G. and De Sarlo, L. and Catani, J. and Minardi, F. and Inguscio, M.},
  journal = {Phys. Rev. Lett.},
  volume = {100},
  issue = {21},
  pages = {210402},
  numpages = {4},
  year = {2008},
  month = {May},
  publisher = {American Physical Society},
  doi = {10.1103/PhysRevLett.100.210402},
  url = {https://link.aps.org/doi/10.1103/PhysRevLett.100.210402}
}

@article{MorizotPRA2006,
  title = {Ring trap for ultracold atoms},
  author = {Morizot, Olivier and Colombe, Yves and Lorent, Vincent and Perrin, H\'el\`ene and Garraway, Barry M.},
  journal = {Phys. Rev. A},
  volume = {74},
  issue = {2},
  pages = {023617},
  numpages = {10},
  year = {2006},
  month = {Aug},
  publisher = {American Physical Society},
  doi = {10.1103/PhysRevA.74.023617},
  url = {https://link.aps.org/doi/10.1103/PhysRevA.74.023617}
}

@article{KSgan2025josephson2D,
  title = {Josephson dynamics in two-dimensional ring-shaped condensates},
  author = {Gan, Koon Siang and Singh, Vijay Pal and Amico, Luigi and Dumke, Rainer},
  journal = {Phys. Rev. Res.},
  volume = {8},
  issue = {2},
  pages = {023190},
  numpages = {8},
  year = {2026},
  month = {May},
  publisher = {American Physical Society},
  doi = {10.1103/3249-x994},
  url = {https://link.aps.org/doi/10.1103/3249-x994}
}

@article{Amico2021roadmap,
    author = {Amico, L. and Boshier, M. and Birkl, G. and Minguzzi, A. and Miniatura, C. and Kwek, L.-C. and Aghamalyan, D. and Ahufinger, V. and Anderson, D. and Andrei, N. and Arnold, A. S. and Baker, M. and Bell, T. A. and Bland, T. and Brantut, J. P. and Cassettari, D. and Chetcuti, W. J. and Chevy, F. and Citro, R. and De Palo, S. and Dumke, R. and Edwards, M. and Folman, R. and Fortagh, J. and Gardiner, S. A. and Garraway, B. M. and Gauthier, G. and Günther, A. and Haug, T. and Hufnagel, C. and Keil, M. and Ireland, P. and Lebrat, M. and Li, W. and Longchambon, L. and Mompart, J. and Morsch, O. and Naldesi, P. and Neely, T. W. and Olshanii, M. and Orignac, E. and Pandey, S. and Pérez-Obiol, A. and Perrin, H. and Piroli, L. and Polo, J. and Pritchard, A. L. and Proukakis, N. P. and Rylands, C. and Rubinsztein-Dunlop, H. and Scazza, F. and Stringari, S. and Tosto, F. and Trombettoni, A. and Victorin, N. and Klitzing, W. von and Wilkowski, D. and Xhani, K. and Yakimenko, A.},
    title = {Roadmap on Atomtronics: State of the art and perspective},
    journal = {AVS Quantum Science},
    volume = {3},
    number = {3},
    pages = {039201},
    year = {2021},
    month = {08},
    issn = {2639-0213},
    doi = {10.1116/5.0026178},
    url = {https://doi.org/10.1116/5.0026178}
}

@article{BlochoscillationsRabec2025,
  author  = {Rabec, F. and Chauveau, G. and Brochier, G. and Nascimbene, S. and Dalibard, J. and Beugnon, J.},
  title   = {Bloch oscillations of a soliton in a one-dimensional quantum fluid},
  journal = {Nature Physics},
  year    = {2025},
  volume  = {21},
  number  = {10},
  pages   = {1541--1547},
  month   = oct,
  doi     = {10.1038/s41567-025-02970-1},
  url     = {https://doi.org/10.1038/s41567-025-02970-1}
}

@misc{Halltori,
      title={Realization of a Synthetic Hall Torus with a Spinor Bose-Einstein Condensate}, 
      author={T. -H. Chien and S. -C. Wu and Y. -H. Su and L. -R. Liu and N. -C. Chiu and M. Sarkar and Q. Zhou and Y. -J. Lin},
      year={2026},
      eprint={2602.14549},
      archivePrefix={arXiv},
      primaryClass={cond-mat.quant-gas},
      url={https://arxiv.org/abs/2602.14549}, 
}

@article{SaitoPhysRevA.82.013647,
  title = {Connection between rotation and miscibility in a two-component Bose-Einstein condensate},
  author = {Shimodaira, Takayuki and Kishimoto, Tetsuo and Saito, Hiroki},
  journal = {Phys. Rev. A},
  volume = {82},
  issue = {1},
  pages = {013647},
  numpages = {6},
  year = {2010},
  month = {Jul},
  publisher = {American Physical Society},
  doi = {10.1103/PhysRevA.82.013647},
  url = {https://link.aps.org/doi/10.1103/PhysRevA.82.013647}
}

@article{NaliniPhysRevResearch.7.023051,
  title = {Signature of Andreev-Bashkin superfluid drag from cavity optomechanics},
  author = {Pradhan, Nalinikanta and Kanamoto, Rina and Bhattacharya, M. and Mishra, Pankaj Kumar},
  journal = {Phys. Rev. Res.},
  volume = {7},
  issue = {2},
  pages = {023051},
  numpages = {11},
  year = {2025},
  month = {Apr},
  publisher = {American Physical Society},
  doi = {10.1103/PhysRevResearch.7.023051},
  url = {https://link.aps.org/doi/10.1103/PhysRevResearch.7.023051}
}

@article{InterferometryPhysRevLett.120.063201,
  title = {Spin-Orbit-Coupled Interferometry with Ring-Trapped Bose-Einstein Condensates},
  author = {Helm, J. L. and Billam, T. P. and Rakonjac, A. and Cornish, S. L. and Gardiner, S. A.},
  journal = {Phys. Rev. Lett.},
  volume = {120},
  issue = {6},
  pages = {063201},
  numpages = {6},
  year = {2018},
  month = {Feb},
  publisher = {American Physical Society},
  doi = {10.1103/PhysRevLett.120.063201},
  url = {https://link.aps.org/doi/10.1103/PhysRevLett.120.063201}
}

@article{RotationPhysRevA.93.023616,
  title = {Quantum enhanced measurement of rotations with a spin-1 Bose-Einstein condensate in a ring trap},
  author = {Nolan, Samuel P. and Sabbatini, Jacopo and Bromley, Michael W. J. and Davis, Matthew J. and Haine, Simon A.},
  journal = {Phys. Rev. A},
  volume = {93},
  issue = {2},
  pages = {023616},
  numpages = {13},
  year = {2016},
  month = {Feb},
  publisher = {American Physical Society},
  doi = {10.1103/PhysRevA.93.023616},
  url = {https://link.aps.org/doi/10.1103/PhysRevA.93.023616}
}

@article{RotaionPhysRevA.81.061602,
  title = {Rotational response of two-component Bose-Einstein condensates in ring traps},
  author = {Halkyard, P. L. and Jones, M. P. A. and Gardiner, S. A.},
  journal = {Phys. Rev. A},
  volume = {81},
  issue = {6},
  pages = {061602(R)},
  numpages = {4},
  year = {2010},
  month = {Jun},
  publisher = {American Physical Society},
  doi = {10.1103/PhysRevA.81.061602},
  url = {https://link.aps.org/doi/10.1103/PhysRevA.81.061602}
}

@article{MassiveexcitationRabiPhysRevLett.128.210401,
  title = {Observation of Massless and Massive Collective Excitations with Faraday Patterns in a Two-Component Superfluid},
  author = {Cominotti, R. and Berti, A. and Farolfi, A. and Zenesini, A. and Lamporesi, G. and Carusotto, I. and Recati, A. and Ferrari, G.},
  journal = {Phys. Rev. Lett.},
  volume = {128},
  issue = {21},
  pages = {210401},
  numpages = {6},
  year = {2022},
  month = {May},
  publisher = {American Physical Society},
  doi = {10.1103/PhysRevLett.128.210401},
  url = {https://link.aps.org/doi/10.1103/PhysRevLett.128.210401}
}

@article{SpinorPersistentStabilityPhysRevA.88.051602,
  title = {Stability of persistent currents in spinor Bose-Einstein condensates},
  author = {Yakimenko, A. I. and Isaieva, K. O. and Vilchinskii, S. I. and Weyrauch, M.},
  journal = {Phys. Rev. A},
  volume = {88},
  issue = {5},
  pages = {051602(R)},
  numpages = {5},
  year = {2013},
  month = {Nov},
  publisher = {American Physical Society},
  doi = {10.1103/PhysRevA.88.051602},
  url = {https://link.aps.org/doi/10.1103/PhysRevA.88.051602}
}

@article{Mathey2016Realizing,
doi = {10.1088/1367-2630/18/5/055016},
url = {https://doi.org/10.1088/1367-2630/18/5/055016},
year = {2016},
month = {may},
publisher = {IOP Publishing},
volume = {18},
number = {5},
pages = {055016},
author = {Mathey, Amy C and Mathey, L},
title = {Realizing and optimizing an atomtronic SQUID},
journal = {New Journal of Physics}
}

@article{Kunimi2019Decay,
  title = {Decay mechanisms of superflow of Bose-Einstein condensates in ring traps},
  author = {Kunimi, Masaya and Danshita, Ippei},
  journal = {Phys. Rev. A},
  volume = {99},
  issue = {4},
  pages = {043613},
  numpages = {9},
  year = {2019},
  month = {Apr},
  publisher = {American Physical Society},
  doi = {10.1103/PhysRevA.99.043613},
  url = {https://link.aps.org/doi/10.1103/PhysRevA.99.043613}
}

@article{kim2025josephson,
  title={Josephson junctions in the age of quantum discovery},
  author={Kim, Hyunseong and Jang, Gyunghyun and Jin, Seungwon and Shin, Dongbin and Shin, Hyeon-Jin and Luo, Jie and Siddiqi, Irfan and Kim, Yosep and Yoon, Hoon Hahn and Nguyen, Long B},
  journal={arXiv preprint arXiv:2505.12724},
  year={2025}
}

@article{kjaergaard2020superconducting,
  title={Superconducting qubits: Current state of play},
  author={Kjaergaard, Morten and Schwartz, Mollie E and Braum{\"u}ller, Jochen and Krantz, Philip and Wang, Joel I-J and Gustavsson, Simon and Oliver, William D},
  journal={Annual Review of Condensed Matter Physics},
  volume={11},
  number={1},
  pages={369--395},
  year={2020},
  publisher={Annual Reviews}
}

@article{POLO20251,
title = {Persistent currents in ultracold gases},
journal = {Physics Reports},
volume = {1137},
pages = {1-70},
year = {2025},
note = {Persistent currents in ultracold gases},
issn = {0370-1573},
doi = {https://doi.org/10.1016/j.physrep.2025.06.003},
url = {https://www.sciencedirect.com/science/article/pii/S0370157325001796},
author = {J. Polo and W.J. Chetcuti and T. Haug and A. Minguzzi and K. Wright and L. Amico}
}

@article{nesti2026increasing,
  title={Increasing the stability of a superfluid in a rotating necklace potential},
  author={Nesti, Giulio and Pezz{\`e}, Luca},
  journal={arXiv preprint arXiv:2601.15159},
  year={2026}
}

@article{ciszak2026cooperative,
  title={Cooperative stabilization of persistent currents in superfluid ring networks},
  author={Ciszak, Marzena and Grani, Nicola and Hernandez-Rajkov, Diego and Del Pace, Giulia and Roati, Giacomo and Marino, Francesco},
  journal={arXiv preprint arXiv:2601.15121},
  year={2026}
}

@article{Gallemi2016Coherent,
doi = {10.1088/1367-2630/18/1/015003},
url = {https://doi.org/10.1088/1367-2630/18/1/015003},
year = {2015},
month = {dec},
publisher = {IOP Publishing},
volume = {18},
number = {1},
pages = {015003},
author = {Gallemí, A and Mateo, A Muñoz and Mayol, R and Guilleumas, M},
title = {Coherent quantum phase slip in two-component bosonic atomtronic circuits},
journal = {New Journal of Physics},

}

@article{PhysRevA.100.013630,
  title = {Transverse instability and disintegration of a domain wall of a relative phase in coherently coupled two-component Bose-Einstein condensates},
  author = {Ihara, Kousuke and Kasamatsu, Kenichi},
  journal = {Phys. Rev. A},
  volume = {100},
  issue = {1},
  pages = {013630},
  numpages = {8},
  year = {2019},
  month = {Jul},
  publisher = {American Physical Society},
  doi = {10.1103/PhysRevA.100.013630},
  url = {https://link.aps.org/doi/10.1103/PhysRevA.100.013630}
}

@article{PhysRevA.80.053602,
  title = {Critical velocity of superfluid flow through single-barrier and periodic potentials},
  author = {Watanabe, Gentaro and Dalfovo, F. and Piazza, F. and Pitaevskii, L. P. and Stringari, S.},
  journal = {Phys. Rev. A},
  volume = {80},
  issue = {5},
  pages = {053602},
  numpages = {9},
  year = {2009},
  month = {Nov},
  publisher = {American Physical Society},
  doi = {10.1103/PhysRevA.80.053602},
  url = {https://link.aps.org/doi/10.1103/PhysRevA.80.053602}
}

@article{PhysRevA.91.063635,
  title = {Rabi-coupled countersuperflow in binary Bose-Einstein condensates},
  author = {Usui, Ayaka and Takeuchi, Hiromitsu},
  journal = {Phys. Rev. A},
  volume = {91},
  issue = {6},
  pages = {063635},
  numpages = {8},
  year = {2015},
  month = {Jun},
  publisher = {American Physical Society},
  doi = {10.1103/PhysRevA.91.063635},
  url = {https://link.aps.org/doi/10.1103/PhysRevA.91.063635}
}

@article{abad2014Persistent,
  title = {Persistent currents in two-component condensates in a toroidal trap},
  author = {Abad, M. and Sartori, A. and Finazzi, S. and Recati, A.},
  journal = {Phys. Rev. A},
  volume = {89},
  issue = {5},
  pages = {053602},
  numpages = {9},
  year = {2014},
  month = {May},
  publisher = {American Physical Society},
  doi = {10.1103/PhysRevA.89.053602},
  url = {https://link.aps.org/doi/10.1103/PhysRevA.89.053602}
}

@article{SpinorPersistentPhysRevLett.110.025301,
  title = {Persistent Currents in Spinor Condensates},
  author = {Beattie, Scott and Moulder, Stuart and Fletcher, Richard J. and Hadzibabic, Zoran},
  journal = {Phys. Rev. Lett.},
  volume = {110},
  issue = {2},
  pages = {025301},
  numpages = {5},
  year = {2013},
  month = {Jan},
  publisher = {American Physical Society},
  doi = {10.1103/PhysRevLett.110.025301},
  url = {https://link.aps.org/doi/10.1103/PhysRevLett.110.025301}
}

@article{PhysRevA.101.033610,
  title = {Effects of atom numbers on the miscibility-immiscibility transition of a binary Bose-Einstein condensate},
  author = {Wen, Lin and Guo, Hui and Wang, Ya-Jun and Hu, Ai-Yuan and Saito, Hiroki and Dai, Chao-Qing and Zhang, Xiao-Fei},
  journal = {Phys. Rev. A},
  volume = {101},
  issue = {3},
  pages = {033610},
  numpages = {9},
  year = {2020},
  month = {Mar},
  publisher = {American Physical Society},
  doi = {10.1103/PhysRevA.101.033610},
  url = {https://link.aps.org/doi/10.1103/PhysRevA.101.033610}
}

@article{Manakov:1974,
 title = {On the theory of two-dimensional stationary
self-focusing of electromagnetic waves},
 author = {Manakov, S. V.},
 journal = {Sov. Phys. JETP},
 volume = {65},
 pages = {248},
 year = {1974}
 }

@article{Sarkar:2025,
	author = {Sarkar, Swarup K and Mardonov, Sh and Ya Sherman, E and Muruganandam, Paulsamy and Mishra, Pankaj K},
	doi = {10.1088/1367-2630/adafd8},
	journal = {New Journal of Physics},
	month = {feb},
	number = {2},
	pages = {023018},
	publisher = {IOP Publishing},
	title = {Spin-dependent localization of spin--orbit and {Rabi}-coupled {Bose-Einstein} condensates in a random potential},
	url = {https://dx.doi.org/10.1088/1367-2630/adafd8},
	volume = {27},
	year = {2025},
}

@article{StringariPhysRevLett.116.160402,
  title = {Magnetic Solitons in a Binary {Bose-Einstein} Condensate},
  author = {Qu, Chunlei and Pitaevskii, Lev P. and Stringari, Sandro},
  journal = {Phys. Rev. Lett.},
  volume = {116},
  issue = {16},
  pages = {160402},
  numpages = {5},
  year = {2016},
  month = {Apr},
  publisher = {American Physical Society},
  doi = {10.1103/PhysRevLett.116.160402},
  url = {https://link.aps.org/doi/10.1103/PhysRevLett.116.160402}
}

@article{FerrariPhysRevLett.125.030401,
  title = {Observation of Magnetic Solitons in Two-Component Bose-Einstein Condensates},
  author = {Farolfi, A. and Trypogeorgos, D. and Mordini, C. and Lamporesi, G. and Ferrari, G.},
  journal = {Phys. Rev. Lett.},
  volume = {125},
  issue = {3},
  pages = {030401},
  numpages = {5},
  year = {2020},
  month = {Jul},
  publisher = {American Physical Society},
  doi = {10.1103/PhysRevLett.125.030401},
  url = {https://link.aps.org/doi/10.1103/PhysRevLett.125.030401}
}

@article{liu2015new,
  title={A new type of half-quantum circulation in a macroscopic polariton spinor ring condensate},
  author={Liu, Gangqiang and Snoke, David W and Daley, Andrew and Pfeiffer, Loren N and West, Ken},
  journal={Proceedings of the National Academy of Sciences},
  volume={112},
  number={9},
  pages={2676--2681},
  year={2015},
  publisher={National Academy of Sciences}
}

@article{Pradhan2026Proposals,
  title = {Proposals for realizing a Josephson diode in atomtronic circuits},
  author = {Pradhan, Nalinikanta and Kanamoto, Rina and Bhattacharya, M. and Mishra, Pankaj Kumar},
  journal = {Phys. Rev. A},
  volume = {113},
  issue = {6},
  pages = {L061502},
  numpages = {7},
  year = {2026},
  month = {Jun},
  publisher = {American Physical Society},
  doi = {10.1103/7kb2-118m},
  url = {https://link.aps.org/doi/10.1103/7kb2-118m}
}

@article{Mathey2014Decay,
  title = {Decay of a superfluid current of ultracold atoms in a toroidal trap},
  author = {Mathey, Amy C. and Clark, Charles W. and Mathey, L.},
  journal = {Phys. Rev. A},
  volume = {90},
  issue = {2},
  pages = {023604},
  numpages = {14},
  year = {2014},
  month = {Aug},
  publisher = {American Physical Society},
  doi = {10.1103/PhysRevA.90.023604},
  url = {https://link.aps.org/doi/10.1103/PhysRevA.90.023604}
}

@article{Kim2020Observation,
  title = {Observation of two sound modes in a binary superfluid gas},
  author = {Kim, Joon Hyun and Hong, Deokhwa and Shin, Y.},
  journal = {Phys. Rev. A},
  volume = {101},
  issue = {6},
  pages = {061601(R)},
  numpages = {6},
  year = {2020},
  month = {Jun},
  publisher = {American Physical Society},
  doi = {10.1103/PhysRevA.101.061601},
  url = {https://link.aps.org/doi/10.1103/PhysRevA.101.061601}
}

@misc{supplement,
  title = {See Supplemental Material at [URL] for additional derivations and results.},
  year = {2026},
}

@article{2D_PhysRevA.111.043308,
  title = {Acceleration-driven dynamics of Josephson vortices in coplanar superfluid rings},
  author = {Borysenko, Yurii and Bazhan, Nataliia and Prykhodko, Olena and Pfeiffer, Dominik and Lind, Ludwig and Birkl, Gerhard and Yakimenko, Alexander},
  journal = {Phys. Rev. A},
  volume = {111},
  issue = {4},
  pages = {043308},
  numpages = {10},
  year = {2025},
  month = {Apr},
  publisher = {American Physical Society},
  doi = {10.1103/PhysRevA.111.043308},
  url = {https://link.aps.org/doi/10.1103/PhysRevA.111.043308}
}


\clearpage

\widetext

\begin{center}
\textbf{\large Supplementary material: Spinor condensate persistent currents in an atomtronic Josephson necklace}
\end{center}

\setcounter{equation}{0} \setcounter{figure}{0} \setcounter{table}{0} %
\setcounter{page}{1} \setcounter{section}{0} \makeatletter
\renewcommand{\theequation}{\arabic{equation}} \renewcommand{\thefigure}{S%
\arabic{figure}} \renewcommand{\bibnumfmt}[1]{[]} \renewcommand{%
\citenumfont}[1]{#1} \renewcommand{\thesection}{\arabic{section}}%
\setcounter{secnumdepth}{3}

\renewcommand{\thefigure}{\arabic{figure}} \renewcommand{\thesection}{S\arabic{section}} \renewcommand{\theequation}{S\arabic{equation}}

In this Supplemental Material we provide the additional theoretical details and numerical results supporting the main text. Section~\ref{sec:SM1} presents the dimensional reduction procedure and the corresponding energy scaling leading to the effective one-dimensional Gross--Pitaevskii equations. 
In Sec.~\ref{SM:polarization} we present the effect of polarization on the stability of the spinor persistent current. In Sec.~\ref{sec:SM4}, we investigate the dynamics and stability of counter-rotating binary persistent currents. Section~\ref{sec:SM5} compares the mean-field Gross--Pitaevskii dynamics with simulations based on the stochastic Gross--Pitaevskii equation, highlighting the robustness of our results against thermal fluctuations. Section~\ref{sec:landau} derives the stability condition from the Landau criterion. In Sec.~\ref{sec:magneticsoliton} we present a detailed analysis of the magnetic soliton which has been observed during the phase slip for higher interspecies interaction. Finally, we discuss the robustness of interspecies interaction in stabilizing the spinor current in two dimension in Sec.~\ref{sec:2D}.
\section{Theoretical framework of mean-field model of ring spinor condensate}
\label{sec:SM1}


We consider a two-component atomic Bose-Einstein Condensate (BEC) confined by a ring-shaped potential, expressed as 
\begin{equation}
    U(\rho, z) = \frac{1}{2}\, m_\sigma \omega_\rho^2 (\rho-R)^2 + \frac{1}{2}\, m_\sigma \omega_z^2 z^2,
\end{equation}
where $\sigma \in \{1,2\}$, $m_1$ and $m_2$ are mass of each atoms of the two components, $R$ is the radius of the ring trap, $\omega_\rho$ is the trapping frequency in the radial direction, and $\omega_z$ is the trapping frequency in the axial direction. We assume that the trapping frequencies are the same for both components. For a tight confinement in the radial and axial directions, we assume that the condensate ground states are frozen in those directions and that dynamics in only the azimuthal direction $\phi$ are relevant. This assumption is justified when the chemical potential of each component $\mu_\sigma < \hbar \omega_\rho,\hbar \omega_z$ and the atom number \cite{MorizotPRA2006, KumarPRL2021} 
\begin{equation}
    N_\sigma\lesssim \frac{4 \sqrt{\pi}R}{3a^{\sigma}_{\mathrm{s}}} \bigg(\frac{\omega_\rho}{\omega_z}\bigg)^{\frac{1}{2}},
\end{equation}
where $a^{\sigma}_{\mathrm{s}}$ is the s-wave scattering length of the two components. With these assumptions, the dynamics of the system of consideration can be explained by a one-dimensional Hamiltonian
\begin{align}
    \hat{H} &= \sum_{\sigma,\sigma'} \; \int_{-\pi}^{\pi} \Bigl\{  \Psi_\sigma^{\dagger} (\phi) \left[\frac{\hbar^2}{2m R^2}\left(-i\frac{d}{d\phi}\right)^2 + V(\phi) \right]\Psi_\sigma(\phi)  \nonumber\\ & 
    +  \frac{{\mathcal{G}_{\sigma \sigma}}}{2}\; \Psi_\sigma^{\dagger}(\phi)\Psi_\sigma^{\dagger}(\phi)  \Psi_\sigma(\phi)\Psi_\sigma(\phi) +  \frac{{\mathcal{G}_{\sigma {\sigma'}}}}{2}\; \Psi_\sigma^{\dagger}(\phi)\Psi_{\sigma'}^{\dagger}(\phi)\Psi_{\sigma'}(\phi)\Psi_\sigma(\phi) \Bigr\} \; d\phi  ,\label{Eq:Hamil}
\end{align}
where $(\sigma,{\sigma'}) \in {1,2}, \sigma\neq {\sigma'}$, $\Psi_\sigma^{\dagger} (\phi)$ and $\Psi_\sigma (\phi)$ are the bosonic creation and annihilation operators, respectively, with commutation $[\Psi_\sigma (\phi),\Psi_\sigma^{\dagger} (\phi')] = \delta(\phi-\phi')$. Here we have the considered the masses of the atoms in the condensates to be equal, i.e., $m_1 = m_2 = m$. The first and second terms in Eq.~\ref{Eq:Hamil} represent the rotational kinetic energy of the condensates and the optical barrier potentials used to model the Josephson junction necklace. The third and fourth terms in Eq.~\ref{Eq:Hamil} represent the one-dimensional (1D) intra-component and inter-component density-density interactions with strength ${\mathcal{G}_{\sigma \sigma}}$ and ${\mathcal{G}_{\sigma \sigma'}}$, respectively. The 1D interaction strengths are reduced from their 3D counterparts, such as ${\mathcal{G}_{\sigma \sigma}} = {\mathcal{G}^{3D}_{\sigma \sigma}} / (2 \pi l_z l_\rho) = 2 \hbar a^{\sigma \sigma}_s \sqrt{\omega_\rho \omega_z}$, where  $l_z = \sqrt{\hbar/(m \omega_z)}$ and $l_\rho = \sqrt{\hbar/(m \omega_\rho)}$ are the harmonic oscillator lengths in the axial and radial directions, respectively and $a^{\sigma \sigma}_s$ is the 3D s-wave scattering length of the two components~\cite{MorizotPRA2006,PoloPRL2019}. By taking into account experimentally feasible parameters such as $a_s = 2.5 $ nm for $^{23}$Na  atom, total number of atoms $N = 10000$ in a ring trap of radius $R = 25 \mu$m, and temperature $T = 10$nK, we have calculated the reduced interaction strength~\cite{PoloPRL2019}
\begin{equation}
    \gamma=\frac{m\mathcal{G_{\sigma \sigma}}}{\hbar^{2}n} = 7.5\times 10^{-6},
\end{equation}
and the reduced temperature
\begin{equation}
    T_r = \frac{T }{T_d} = 0.00024,
\end{equation}
where $T_d = \hbar^2 n^2/2m k_B$ is the condensation temperature and $n = N/(2\pi R)$ is the condensate density. These values ($\gamma < 1$ and $T_r \approx 0$) allow us to justify a 1D mean-field approach using the Gross-Pitaevskii equation (GPE) for the Hamiltonian (Eq.~\ref{Eq:Hamil}) as
\begin{equation}
\begin{aligned}
    i\hbar\frac{d \psi_{\sigma}(\phi, t)}{dt} = -\frac{\hbar^2}{2m R^2}\frac{d^2 \psi_{\sigma}(\phi, t)}{d\phi ^2} + V(\phi, t) + (\mathcal{G}_{\sigma \sigma}  |\psi_{\sigma}|^2 + \mathcal{G}_{\sigma {\sigma'}}  |\psi_{\sigma'}|^2 ) \, \psi_{\sigma}(\phi, t).
    \label{Eq:bec}    
\end{aligned}
\end{equation}
After scaling the energy as $\hbar \omega_\beta = \frac{\hbar^2}{2m R^2} $, time as $\tau = \omega_\beta t$, we arrive at the dimensionless 1D Gross-Pitaevskii equation
\begin{equation}
\begin{aligned}
    i\frac{d \psi_{\sigma}(\phi, \tau)}{d\tau} =  -\frac{d^2 \psi_{\sigma}(\phi, \tau)}{d\phi ^2} + \tilde{V}(\phi, \tau) + (G_{\sigma \sigma}  |\psi_{\sigma}|^2 + G_{\sigma {\sigma'}}  |\psi_{\sigma'}|^2 ) \, \psi_{\sigma}(\phi, \tau)  ,
    \label{Eq:bec_scaled}    
\end{aligned}
\end{equation}
where $\tilde{V}(\phi, \tau) = V(\phi, \tau)/\hbar \omega_\beta$, $G = \mathcal{G} / \hbar \omega_\beta$ is the scaled 1D interaction strength. For all the results provided in the manuscript except the coherently coupled condensates cases, we have simulated Eqs.~\ref{Eq:bec_scaled} up to $0.5$ seconds, which matches the time scale of previous experiments on persistent current decay \cite{Luca2024Stabilizing}.

\section{Effect of Polarization}
\label{SM:polarization}
In addition to the barrier parameters and intercomponent interactions, the population imbalance between the two components provides another effective control parameter for the stability of spinor persistent currents. We define the polarization as $P=(N_1-N_2)/(N_1+N_2)$, where $N_1$ and $N_2$ are the particle numbers in the two components~\cite{SpinorPersistentPhysRevLett.110.025301}. The condensate is prepared with unequal circulation states, $L_{p1}=1$ and $L_{p2}=0$, and the stability of the rotating component is examined as a function of $P$.
\begin{figure}[!htb]
\centering
\includegraphics[width= 0.5\linewidth]{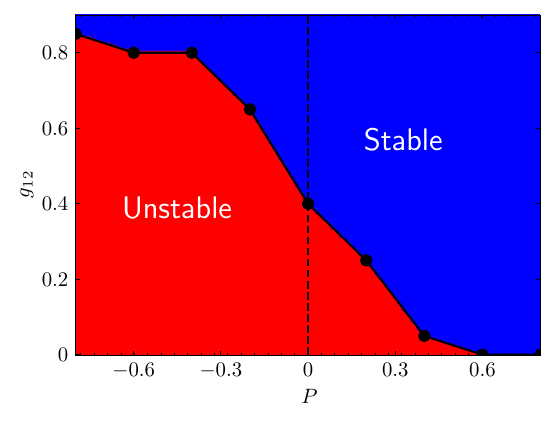}
\caption{Stability phase diagram in the interspecies interaction $g_{12}$ and polarization $P$ parameter space. The red region indicates the unstable region, while the blue region denotes the stable one. Here $g_{11} = g_{22}$ \cite{PhysRevA.101.033610} and the other set of parameters used here is the same as used in Fig. 1. }
\label{fig:pol_g12}
\end{figure}
As can be seen from Fig.~\ref{fig:pol_g12}, the persistent current is strongly polarization- dependent. For positive polarization, the rotating component contains more atoms, leading to weaker density depletion at the Josephson barriers. This increases the energy required for soliton nucleation, suppresses phase slips, and stabilizes the superflow. In contrast, negative polarization enhances the density depletion at the barriers, facilitating soliton emission and promoting persistent-current decay. Consequently, the system is unstable for sufficiently negative $P$ but becomes increasingly stable as the polarization is increased.

The critical polarization required for stability can be further reduced by increasing the intercomponent interaction strength $g_{12}$. The corresponding $P-g_{12}$ phase diagram in Fig.~\ref{fig:pol_g12} shows that the stable region expands with both increasing polarization and stronger intercomponent coupling. Polarization and intercomponent interactions therefore provide complementary and experimentally accessible knobs for controlling phase-slip dynamics and stabilizing quantized supercurrents in spinor Bose--Einstein condensates.

\section{Counter-rotating binary persistent currents}
\label{sec:SM4}
\begin{figure}[!htbp]
\includegraphics[width= 0.5\linewidth]{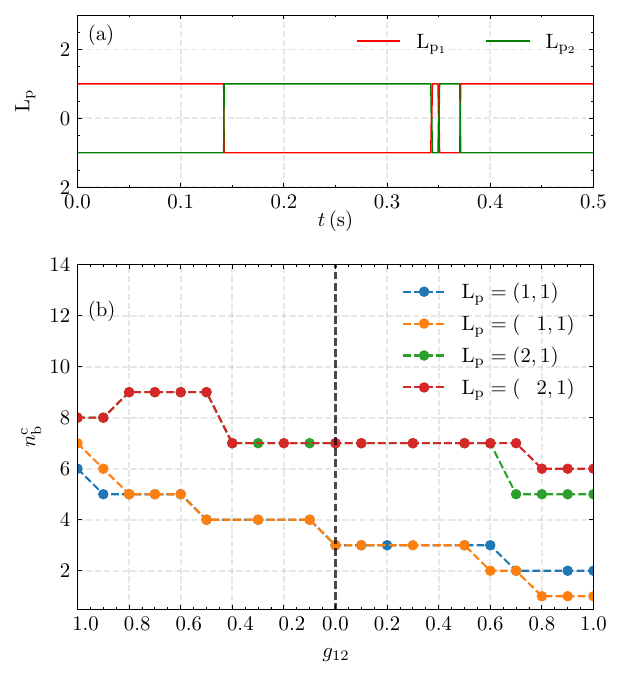}\\
\caption{  (a) Temporal evolution of winding numbers for $L_p = (-1,1)$, $n_b = 2$ and $g_{12} = 0.1$. (b) Critical number of barrier ($n_{b_c}$) required to stabilize co-and counter-rotating persistent current versus density-density interaction strength ($g_{12}$). Here, $V_0 = 3.0 \mu$ and the other set of parameters is the same as used in Fig. 1 of the main manuscript.}
\label{fig:nk_nb_negLp}
\end{figure}

In this section, we compare the stability of co-rotating and counter-rotating binary persistent currents with the same winding-number magnitude, considering $L_p=(1,1)$ and $L_p=(-1,1)$. The corresponding critical number of barriers, $n_{b_c}$, as a function of the interspecies interaction strength $g_{12}$ is shown in Fig.~\ref{fig:nk_nb_negLp}. For an isotropic ring, the Landau criterion depends on the magnitude of the local superfluid velocity and is invariant under reversal of the flow direction. Consequently, within the present symmetric model, the stability of a persistent current is expected to depend on $|L_p|$ rather than on the sign of the winding number. The density dynamics therefore remain essentially identical for the co-rotating and counter-rotating configurations, yielding nearly the same critical barrier number for $L_p=(1,1)$ and $L_p=(-1,1)$.

A representative temporal evolution of the winding numbers for the counter-rotating state, $L_p=(-1,1)$, with $n_b=2$ and $g_{12}=0.1$ is shown in Fig.~\ref{fig:nk_nb_negLp}. The comparison of the critical barrier number for the two configurations is summarized in Fig.~\ref{fig:nk_nb_negLp}(b), where $n_{b_c}$ is plotted as a function of $g_{12}$. The close agreement between the two curves confirms that reversing the circulation of one component does not significantly alter the phase-slip threshold. A small deviation is nevertheless visible near the miscible--immiscible boundary, where the persistent-current states become increasingly susceptible to spin-density fluctuations. In this regime, the softening of the spin mode makes the dynamics particularly sensitive to small differences in the relative-flow configuration, resulting in a weak departure from the otherwise sign-independent stability criterion.

\section{Finite temperature stability of spinor persistent currents}
\label{sec:SM5}

By taking into account thermal fluctuations and damping of the condensates, the governing dynamical equations for the spinor persistent current are described by the stochastic Gross–Pitaevskii equation (sGPE) \cite{stoof2001dynamics, das2012winding, pradhan2024cavity, pradhan2025AndreevBashkin}. The resulting equations for the two components are

\begin{align}
(i-\Gamma)& \frac{\partial \psi_1(\phi,\tau)}{\partial \tau}
=
\Bigl[-\frac{\partial^2}{\partial \phi^2}+ V_n(\phi)+ G_{11} |\psi_1(\phi,\tau)|^2 \Bigr. \nonumber \\ & \quad \Bigl. + G_{12} |\psi_2(\phi,\tau)|^2 \Bigr]\psi_1(\phi,\tau)
 + \Omega \, \psi_2(\phi,\tau)+ \xi(\phi,\tau),
\label{Eq:spinorBEC1_noise}
\\[1em]
(i-\Gamma)& \frac{\partial \psi_2(\phi,\tau)}{\partial \tau}
=
\Bigl[-\frac{\partial^2}{\partial \phi^2}+ V_n(\phi)+ G_{22} |\psi_2(\phi,\tau)|^2 \Bigr. \nonumber \\ & \quad \Bigl. + G_{12} |\psi_1(\phi,\tau)|^2\Bigr]\psi_2(\phi,\tau)
 + \Omega \, \psi_1(\phi,\tau)+ \xi(\phi,\tau).
\label{Eq:spinorBEC2_noise}
\end{align}
Here, $\Gamma$ is the phenomenological damping parameter, while $\xi(\phi,\tau)$ represents thermal fluctuations and is modeled as 
delta-correlated white noise with 
\begin{align} 
\langle\xi(\phi,\tau) \, \xi ^ *(\phi',\tau')\rangle  &= \frac{2\,\Gamma \,k_B \,T}{\hbar \omega_\beta} \, \delta(\phi - \phi') \, \delta (\tau - \tau'), 
\end{align}
Where, $T$ is the temperature of the condensate, $\xi^{*}$ is the complex conjugate of $\xi(\phi,\tau)$ and  $k_B$ is the Boltzmann constant.
\begin{figure}[!htbp]
\includegraphics[width= 0.5\linewidth]{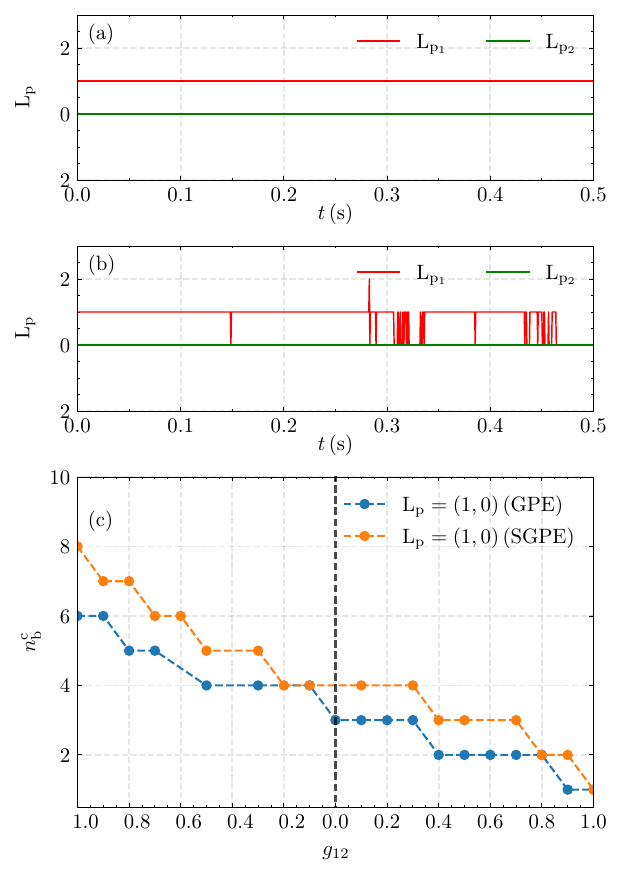}\\
\caption{(a) GPE results and (b) sGPE results ($T = 10$ nK) showing temporal evolution of winding numbers for $n_b = 3$ and $g_{12} = 0.1$.  (c) Comparison between GPE and sGPE results for the critical number of barrier ($n_{b_c}$) required to stabilize the binary-persistent current versus density-density interaction strength ($g_{12}$). Here, $\Gamma = 0.001$, $V_0 = 3.0 \mu$ and the other set of parameters is the same as used in Fig. 1 of the main manuscript.}
  \label{fig:nk_nb_sgpe}
\end{figure}
To assess the robustness of our analysis to thermal fluctuations of the condensate we compare the finite-temperature sGPE (Eqs. \ref{Eq:spinorBEC1_noise} and \ref{Eq:spinorBEC2_noise}) dynamics with the corresponding zero-temperature GPE results presented in the main text. Figures~\ref{fig:nk_nb_sgpe}(a) and (b) show the temporal evolution of the winding numbers obtained from the GPE and sGPE, respectively, for $n_b=3$ and $g_{12}=0.1$. At $T=10nK$ thermal fluctuations continuously perturb the local phase and velocity fields, thereby facilitating the nucleation of phase slips below the deterministic instability threshold. Consequently, the winding number exhibits stronger fluctuations and an earlier loss of circulation in the sGPE dynamics compared with the corresponding GPE evolution.

The effect of thermal fluctuations on the stability threshold is further quantified in Fig.~\ref{fig:nk_nb_sgpe}(c), which compares the critical number of barriers, $n_{b_c}$, required to stabilize the spinor persistent current as a function of $g_{12}$. At finite temperature, a larger number of barriers is generally required to stabilize the current than in the zero-temperature GPE case. This shift reflects the additional phase-slip channel introduced by thermal fluctuations, which lowers the effective stability threshold of the circulating state.
\begin{figure*}[!htbp]

\includegraphics[width= 0.47\linewidth]{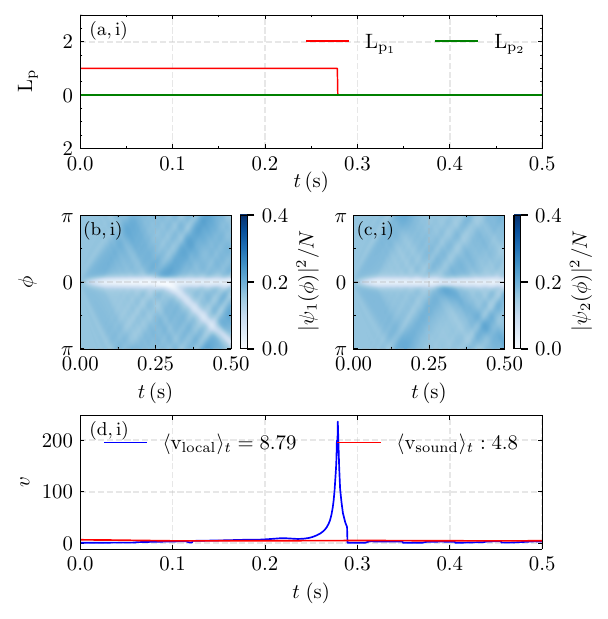}
\includegraphics[width= 0.47\linewidth]{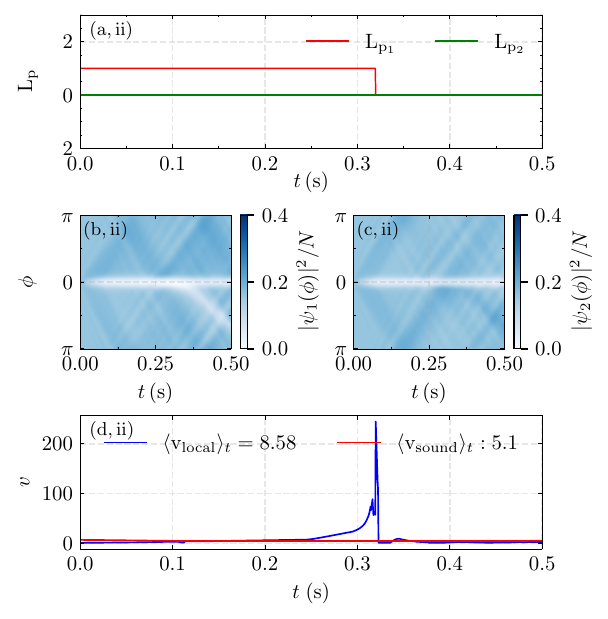}\\
\includegraphics[width= 0.47\linewidth]{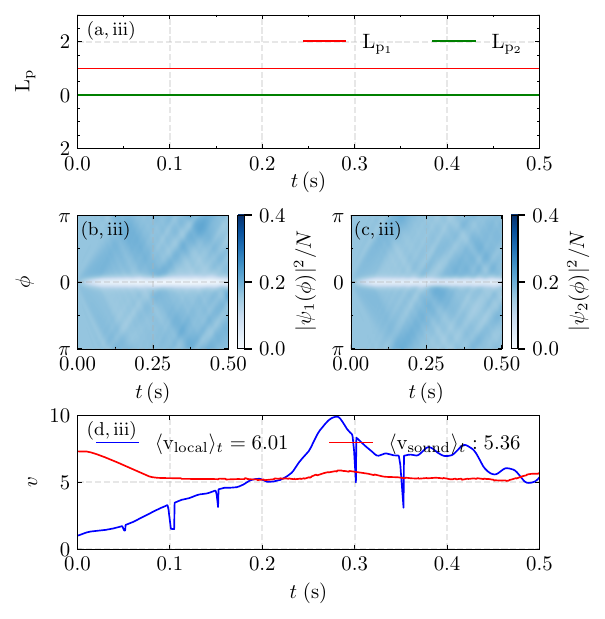}
\includegraphics[width= 0.47\linewidth]{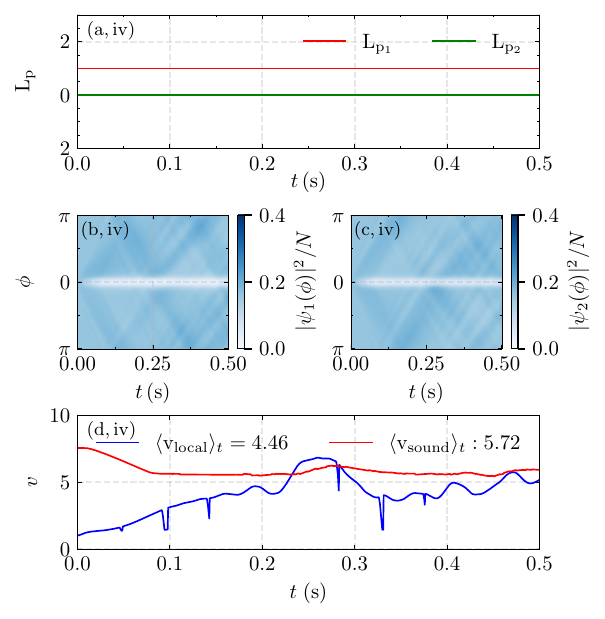}

\caption{  Local Landau criterion for increasing interspecies interaction, $g_{12}=0.2$, $0.3$, $0.4$, and $0.5$, denoted by (i)--(iv), respectively. For each $g_{12}$, panels (a)--(c) show the temporal evolution of the winding number and the densities $|\psi_1|^2$ and $|\psi_2|^2$, respectively, while panel (d) compares the local flow velocity $v_{\mathrm{local}}$ with the density-mode sound velocity $v_{\mathrm{sound}}$ at the barrier. The time-averaged values of the two velocities are indicated in the legends. The parameters are $L_{p_1}=0$, $L_{p_2}=1$, $n_b=1$, and $V_0=2.1\mu$ ($\sigma = 1.2\xi$), with all other parameters as in Fig. 1 of the main manuscript.}
  \label{fig:sound}
\end{figure*}
Importantly, the the qualitative dependence on the interspecies interaction remains unchanged. Increasing $g_{12}$ continues to reduce the critical barrier number in both the GPE and sGPE simulations. The stabilizing effect of interspecies interactions is therefore robust against thermal fluctuations, although the finite-temperature system is quantitatively less stable. Physically, increasing $g_{12}$ enhances the energetic cost of relative-density fluctuations and modifies the density profile around the barriers, reducing the local superfluid velocity and suppressing phase-slip nucleation~\cite{Mathey2014Decay}. Thermal fluctuations act in the opposite direction by perturbing the phase and density fields and providing an additional pathway for phase-slip formation. Thus, thermal noise shifts the stability boundary but does not alter the underlying mechanism responsible for the stabilization of the spinor persistent current.


\section{Condition for stability-Landau criterion}
\label{sec:landau}

The stability of a persistent current can be understood from the Landau criterion, according to which a superflow becomes unstable when its local velocity exceeds the relevant critical excitation velocity~\cite{PhysRevA.80.053602,nesti2026increasing}. In a binary condensate, the elementary collective excitations comprise two branches, namely, a density mode, in which the densities of the two components oscillate in phase, and a spin mode, in which they oscillate out of phase. The corresponding sound velocities provide the natural velocity scales for the onset of flow instabilities. Since the presence of a barrier strongly depletes the condensate density and consequently enhances the local flow velocity, the barrier region becomes the most favorable location for phase-slip nucleation. We therefore examine the local Landau criterion by comparing the flow velocity at the barrier with the density-mode sound velocity.

For the rotating component, $L_{p_2}\ne 0$, the local velocity is determined from the phase gradient as 
\begin{equation}
    v_{\mathrm{local}} = \mathrm{max}[\nabla S],
    \label{eq:v_local}
\end{equation}
where $S = |\mathrm{arg(\psi_2)}|$. The density-mode sound velocity for a binary condensate is given by~\cite{abad2014Persistent} 
\begin{equation}
    v_{\mathrm{sound}} = \sqrt{\frac{g_{11} n^{\mathrm{min}}_1 + g_{22} n^{\mathrm{min}}_2 + \sqrt{(g_{11} n^{\mathrm{min}}_1 - g_{22} n^{\mathrm{min}}_2)^2 + 4 n^{\mathrm{min}}_1 n^{\mathrm{min}}_2 g_{12}^2}  }{2}}.
    \label{eq:v_sound}
\end{equation}

Here, $n^{\mathrm{min}}_1$ and $n^{\mathrm{min}}_2$ are evaluated at a distance of one healing length from the respective density minima at the barrier. We use this offset rather than the density minimum itself to account for the contribution of the bulk sound velocity to the Landau critical velocity~\cite{nesti2026increasing}. The comparison is performed after the barrier ramp has reached its maximum value, and the flow and sound velocities are time averaged over the subsequent evolution.

Figure~\ref{fig:sound} illustrates this Landau-criterion analysis for increasing interspecies interaction, $g_{12}=0.2, 0.3, 0.4,$ and $0.5$. Panels (a)-(c) show the corresponding winding-number and density dynamics of the two components, while panel (d) compares the local flow velocity $v_{local}$ with the density-mode sound velocity $v_{sound}$ at the barrier. For smaller $g_{12}$, the local flow velocity exceeds the density sound velocity, $v_{local}>v_{sound}$, and the persistent current undergoes a phase slip. With increasing $g_{12}$ the interspecies interaction modifies the local density profile and enhances the density-mode sound velocity, while simultaneously suppressing the velocity buildup at the barrier. The two velocity scales consequently approach one another, leading to an enhanced stability of the persistent current. 

The flow velocities exhibit pronounced oscillations even after the barrier reaches its maximum height. These oscillations originate from the dynamical response of the condensate to the barrier ramp and can produce transient crossings of the Landau threshold~\cite{Mathey2014Decay}. We therefore compare the time-averaged values of $v_{local}$ and $v_{sound}$, indicated by the quantities in angular brackets in the legends of Fig.~\ref{fig:sound}(d). The onset of stability occurs when the averaged local flow velocity falls below the corresponding density-mode sound velocity. This provides a direct physical interpretation of the stabilization with increasing $g_{12}$: stronger interspecies interactions increase the restoring energy associated with density fluctuations while suppressing the velocity enhancement at the barrier, thereby moving the system away from the local Landau instability condition. 

\section{Evidence of magnetic solitons for high $g_{12}$}
\label{sec:magneticsoliton}

\begin{figure*}[!htb]
\centering
\includegraphics[width=1.0\linewidth]{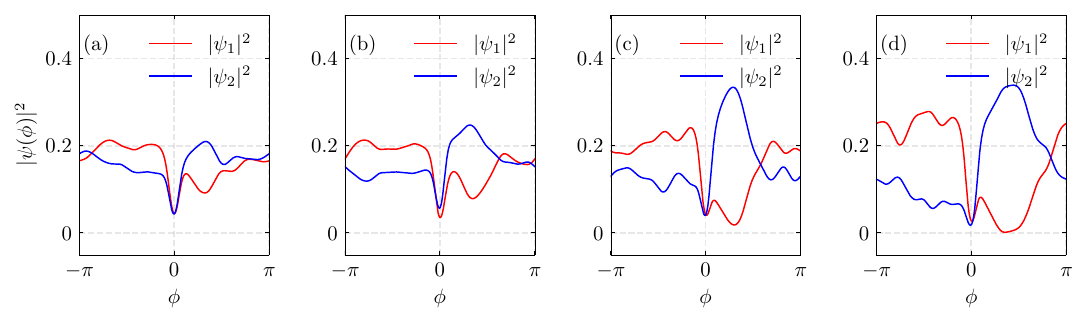}
\caption{Formation of a magnetic soliton for strong interspecies interactions. The soliton width and the associated local magnetization increases markedly as $g_{12}$ approaches the miscible-immiscible threshold, as shown for (a) $g_{12}=0.6$, (b) $0.8$, (c) $0.9$ and (d) $0.96$.}
\label{Mag_sol}
\end{figure*}

To elucidate the enhanced stability observed at strong interspecies interaction [Fig. 2(c) in the main manuscript], we examine the nature of the nonlinear excitations generated by a phase slip as $g_{12}$ approaches the miscible--immiscible threshold. Ground-state solutions are evolved in real time while the Josephson barriers are ramped up during the first $70\,\mathrm{ms}$. Figure~\ref{Mag_sol} shows the corresponding density profiles, $|\psi_i|^2$ ($i=1,2$), for $g_{12}=0.6$, $0.8$, $0.9$ and $0.96$, evaluated shortly after the phase slip. For each value of $g_{12}$, the barrier height is chosen to be at the critical value [determined from Fig.2(c) of the main manuscript] such that the phase slip occurs at $t=0.5\,\mathrm{s}$. This procedure allows us to compare the character of the excitation at a fixed dynamical stage while varying only the interspecies interaction.

For moderate $g_{12}$, a phase slip in the rotating component generates a dark soliton at the barrier, accompanied by a localized density enhancement in the nonrotating component. The density depletion associated with the dark soliton modifies the intercomponent interaction energy and acts as an effective potential for the second component, drawing density into the soliton core. The resulting composite dark--bright structure is therefore characterized by an out-of-phase redistribution of density between the two components. As $g_{12}$ is increased toward the immiscible limit, the spin mode becomes increasingly soft, allowing the relative densities to deform over progressively larger spatial scales. Consequently, the density filling of the soliton core by the nonrotating component becomes more pronounced, producing a strong local polarization, or magnetization, of the two-component fluid. This composite excitation is the characteristic magnetic soliton of a strongly interacting two-component condensate ~\cite{StringariPhysRevLett.116.160402,FerrariPhysRevLett.125.030401}.

The spatial extent of this excitation is governed by the spin healing length, which sets the characteristic length scale for relative-density and spin-texture variations. As $g_{12}\rightarrow1$, the spin mode softens and the spin healing length increases strongly. The magnetic soliton therefore broadens as the miscible--immiscible boundary is approached, as clearly evident in Fig.~\ref{Mag_sol}(a)--(d). The pronounced increase in soliton width provides evidence that the enhanced stability at large $g_{12}$ is accompanied by a qualitative change in the character of the phase-slip excitation, from a conventional density soliton toward an extended magnetic, spin-like excitation.
For completeness we mention that in this limit the density mode becomes stiff as the corresponding healing length becomes very small.

\section{Role of interspecies interaction in stabilizing the spinor current in two-dimensions}
\label{sec:2D}

Here, to examine the robustness of the stabilization mechanism beyond the one-dimensional ring geometry considered in the main manuscript (for reasons of simplicity), we extend our analysis to two dimensions, which is more easily accessible to recent experiments \cite{KSgan2025josephson2D}. We first obtain the ground state by propagating the following coupled Gross--Pitaevskii equations in imaginary time in the presence of a Josephson barrier. Specifically,

\begin{align} 
i \frac{\partial \psi_1(x, y, \tau)}{\partial \tau}
&=
\Bigl[-\nabla^2+ V_t(x,y)+ G_{11} |\psi_1(x, y,\tau)|^2+ G_{12} |\psi_2(x, y,\tau)|^2 \Bigr]\psi_1(x, y,\tau)
 ,
\label{Eq:spinorBEC1_2D}
\\[1em]
i \frac{\partial \psi_2(x,y,\tau)}{\partial \tau}
&=
\Bigl[-\nabla^2+ V_t(x, y)+ G_{22} |\psi_2(x, y,\tau)|^2 + G_{12} |\psi_1(x,y,\tau)|^2\Bigr]\psi_2(x,y,\tau)
 .
\label{Eq:spinorBEC2_2D}
\end{align}
$V_t = V_r + V_b$ represents the total external potential where $V_r$ is the ring--potential that is required to create a toroidal structure of BEC in a flat 2D surface and $V_b$ is the Josephson junction centered at the middle of the annular ring. We model the ring potential as per in \cite{2D_PhysRevA.111.043308} so that we obtain a toroidal condensate with inner radius $R_{in}=12 \mu m$ and outer radius $R_{out} = 24 \mu m$ . The Josephson junction is modelled as a Gaussian barrier by $V_b = V_0 e^{ \frac{(x-x_b)^2}{\delta_x^2} + \frac{(y-y_b)^2}{\delta_y^2} }$. This barrier is centered at $(x_b,y_b)$ with a height of $V_0$. Here, $\nabla^2\equiv \partial_x^2+\partial_y^2$ denotes the two-dimensional Laplacian in polar coordinates. The widths along $x$ and $y$ direction are controlled by the parameters $\delta_x$ and $\delta_y$, which we have considered as $\delta_x=0.5$ and $\delta_x=0.1$ respectively. The resulting state obtained through imaginary time propagation is then evolved in real time after an instantaneous quench of the circulation, implemented through $\psi_\sigma = \psi_\sigma e^{i L_{p\sigma}\phi}$ [with $\sigma\in \{1,2\}$] following the procedure used in the Ref.~\cite{Luca2024Stabilizing}. We consider the space resolution as $\Delta x=\Delta y=0.02$, while, the $d\tau=5\times 10^{-5}$. We impose $L_{p_1}=1$ on the first component while keeping the second component nonrotating, $L_{p_2}=0$, thereby preparing the binary persistent-current state $L_p=(1,0)$.

In two dimensions, the instability develops through vortex nucleation rather than the solitonic excitations characteristic of the one-dimensional geometry. When the local flow at the barrier exceeds the critical velocity, a vortex is nucleated at the barrier and subsequently propagates into the condensate, providing the mechanism for phase-slip-induced decay of the circulating state. Importantly, the interspecies interaction continues to control this instability. Increasing $g_{12}$ progressively suppresses vortex nucleation and can ultimately stabilize the circulating component against phase slips. 

\begin{figure*}[!htb]
\centering
\includegraphics[width=1.0\linewidth]{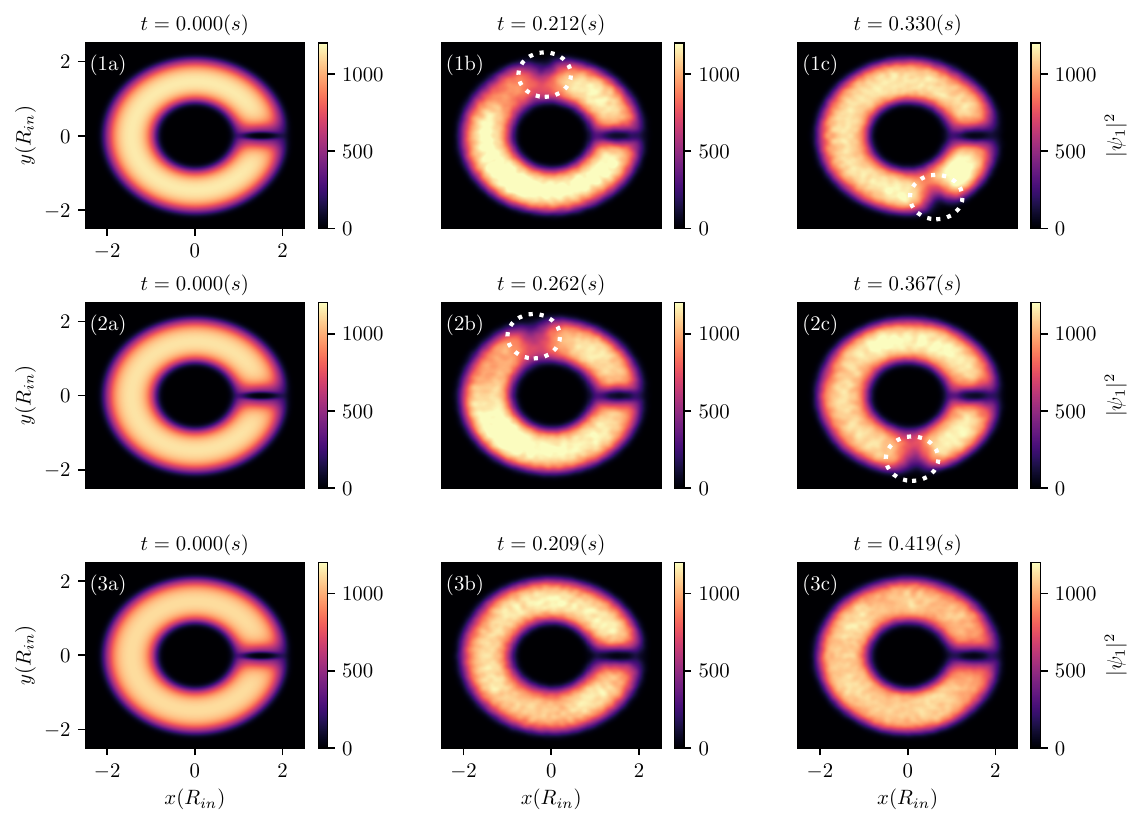}
\caption{ Density evolution of the rotating component, $|\psi_1|^2$, following an instantaneous preparation of the binary persistent current $L_p=(1,0)$ for $g_{12}=0.1$, $0.2$, and $0.3$, shown in rows 1, 2, and 3, respectively. The columns correspond to successive times. For $g_{12}=0.1$ and $0.2$, the circulating state is unstable and decays through vortex nucleation at the barrier, indicated by the dotted white arc. In contrast, the current remains stable for $g_{12}=0.3$, demonstrating the stabilization of the binary persistent current with increasing interspecies interaction. We have taken the number of particles in each component $N_1 = N_2 =10000$. The spatial lengths of $x$ and $y$ are in the units of the inner radius of the ring $(R_{in})$, where, $R_{in}=12\mu m$. The barrier height is fixed to $V_0 = 4.5 \mu$ ~\cite{Luca2024Stabilizing,KSgan2025josephson2D}.}
\label{2D_results}
\end{figure*}

Figure~\ref{2D_results} illustrates this behavior for a single barrier of height $V_0=4.5\mu$. The three rows correspond to $g_{12}=0.1$, $0.2$, and $0.3$, respectively, while the panels within each row show the density evolution of the rotating component, $|\psi_1|^2$, at successive times. For $g_{12}=0.1$ and $0.2$, the $L_{p_1}=1$ state is unstable: a vortex is nucleated at the barrier, indicated by the dotted white arc, and the subsequent vortex motion leads to the decay of the persistent current. In contrast, for $g_{12}=0.3$, no vortex nucleation is observed over the evolution time and the winding remains intact.

Thus, although the nature of the phase-slip excitation changes from soliton emission in one dimension to vortex nucleation in two dimensions, the underlying stabilization mechanism remains the same. The increasing interspecies interaction suppresses the local instability associated with the barrier and prevents the decay of the circulating component. The two-dimensional results therefore provide an independent confirmation of the stabilization mechanism identified in the one-dimensional analysis and demonstrate that the interaction-induced protection of the binary persistent current is not a consequence of the reduced dimensionality. More extensive simulations in two dimensions will be published elsewhere.

\end{document}